\documentstyle[12pt,aaspp4]{article}

\begin{document}
\include{psfig}

\def\Lya{Ly$\alpha$ }
\def\lya{Ly$\alpha$ }
\def\kms{km~s$^{-1}$ }
\def\cm2{\, \rm cm^{-2}}
\def\N#1{{N({\rm #1})}}
\def\rAA{{\rm \, \AA}}
\def\sci#1{{\rm \; \times \; 10^{#1}}}

\title{A KECK HIRES INVESTIGATION OF THE METAL ABUNDANCES AND
KINEMATICS OF THE $z$ = 2.46 DAMPED {\Lya} SYSTEM TOWARD
Q0201+365}

\author{JASON X. PROCHASKA \& ARTHUR M. WOLFE\altaffilmark{1}}
\affil{Department of Physics, and Center for Astrophysics and Space
Sciences; \\
University of California, San
Diego; \\
C--0111; La Jolla; CA 92093\\
}

\altaffiltext{1}{Visiting Astronomer, W.M. Keck Telescope.
The Keck Observatory is a joint facility of the University 
of California and the California Institute of Technology.} 

\begin{abstract} 

        We present high resolution ($\approx 8$ \kms) spectra of the QSO
Q0201+365 obtained with HIRES, the echelle spectrograph 
on the 10m W.M. Keck Telescope.
Although we identify over $80\%$ of the absorption features 
and analyze several of the
more complex metal-line systems,  
we focus our analysis on the damped \Lya system at $z=2.462$.
Ionization simulations suggest the hydrogen in this system 
is significantly neutral and
all of the observed metals are predominantly singly ionized.  
We measure accurate
abundances for Fe, Cr, Si, Ni and place a lower limit on the 
abundance of Zn: 
[Fe/H] = $-0.830 \pm 0.051$, [Cr/H] = $-0.902 \pm 0.064$, 
[Si/H] = $-0.376 \pm 0.052$, [Ni/H] = $-1.002 \pm 0.054$ and 
[Zn/H] $> -0.562 \pm 0.064$.  
We give evidence suggesting the actual Zn abundance
is [Zn/H] $\approx -0.262$, implying the highest 
metallicity observed at a redshift $z \geq 2$.  The
relative abundances of these elements remains constant 
over essentially the entire
system ($\approx 150$ \kms in velocity space), suggesting it is well mixed.
Furthermore, we use the lack of abundance variations to infer properties
of the dust responsible for element depletion.
Finally, we discuss the kinematic characteristics of this
damped \Lya system, comparing and contrasting it with 
other systems.  The low-ion
line profiles span $\approx 200$ \kms in velocity space 
and have an asymmetric shape with
the strongest feature on the red edge.  These kinematic
characteristics are consistent with a rotating disk model.

\end{abstract}


\section{INTRODUCTION}

  This paper is the second in a series devoted to studying the 
metal content of high-redshift galaxies and their progenitors.
Our primary objectives are

\indent (1) to record the emergence of metals in galaxies, 

\indent (2) to trace the mean cosmic metallicity from
$z$ $\approx$ 4.5 to the present,

\indent (3) to determine the kinematic state of galaxies
from $z$ $\approx$ 4.5 to the present. 

\noindent We are implementing this study using HIRES, the echelle
spectrograph on the Keck 10m telescope (\cite{vgt92}), to obtain 
high-resolution spectra of QSOs with foreground damped {\Lya}
systems. The damped {\Lya} systems are a population of neutral
gas layers exhibiting properties indicating they 
are either galaxy progenitors, 
or well formed galaxies detected during an early evolutionary phase. Recent
studies indicate the comoving density of neutral gas in damped systems 
at $z
\approx 3.3$ is comparable to the density of visible 
stars in current galaxies.  At
lower redshifts, the comoving density of neutral gas decreases 
with time in a
manner consistent with gas consumption by star formation 
(\cite{wol95}).
Therefore, studies of the metal content of the damped
\Lya systems enable one to trace the chemical evolution 
of representative galaxies from a
presumably metal-poor gaseous progenitor phase to metal 
rich epochs when most
of the baryons are in stars. As a result, the 
age-metallicity relation, kinematic
conditions, etc., deduced from the damped \Lya systems should tell us more
about the history of galaxies at large 
redshifts than analogous relations deduced
from old stars found in the solar neighborhood (\cite{evd93}).

In a previous paper we presented echelle spectra of the $z$ = 2.309 damped
system toward PHL 957 at a spectral resolution of $\approx$ 8 {\kms} (FWHM)
and signal-to-noise ratio of $\approx$ 35:1 (\cite{wol94}). By fitting
multiple Voigt velocity components to low-ion transitions 
such as Zn II 2026, Ni II
1741, and Cr II 2062 we obtained accurate abundances for 
Zn, Ni, and Cr in the
neutral gas.  The Zn and Cr abundances 
were accurate because we resolved Zn II 
2062.664 from Cr II 2062.234  for the first time in a 
QSO absorption system. We
found that the abundances relative to solar 
were low: [Zn/H] = $-$1.55$\pm$0.11,
[Cr/H] = $-$1.79$\pm$0.10, and [Ni/H] = $-$2.13$\pm$0.08. 
The Zn abundance is especially significant because Zn 
is relatively undepleted by grains in the ISM
of the Galaxy (\cite{sem95}) and is presumably unaffected by dust which
may be present in damped \Lya systems (\cite{fal93}). We also found
the line profiles to be asymmetric in the sense 
that the low column density gas
was found in absorption only at velocities 
higher than the high column-density
gas. The kinematics can be explained by the passage of 
the line of sight through a
rotating disk in which the density of clouds decreases 
with radius and with perpendicular
distance from midplane.

     The purpose of this  paper is to present 
HIRES spectra for Q0201+365, a $V$
= 17.5 QSO with emission redshift $z_{em}$ = 2.49. 
While we identify more than
$80\%$ of the absorption features and find a total of 
13 metal-line redshift
complexes, the focus of this paper is on the damped 
\Lya system at $z$ = 2.462
(\cite{lzwt91,lwt93}).  \cite{lwt93} studied this system at
a resolution of $\approx$ 50 {\kms}. They fitted a 
Voigt damping  profile to
the \Lya absorption trough and found $\N{HI} = 2.4 
\sci{20} \cm2$.  They also identified the metal 
transitions Si II 1190, Si II 1193, Si III 1206, Si II 1260, and
Fe II 1144. Because these transitions are (a) saturated 
and (b) in the \Lya
forest, neither the abundances nor kinematics 
were accurately measured. The
present study represents a major improvement over the 
previous work for the
following reasons. First, we obtain spectra at a 
resolution of $\approx 8$ {\kms}
and a typical signal-to-noise ratio of 33:1. Second, in contrast to 
the previous work,
we focus on metal lines redward of \Lya emission 
where confusion with \Lya
forest lines is absent. Third, because of the higher 
accuracy of the data we focus
on weak unsaturated transitions of ions expected 
to dominate the ionization state
of gas in neutral clouds.  In fact we establish 
accurate element abundances for Fe,
Si, Ni, and Cr, and a lower limit on the abundance of Zn.  
Furthermore, we use computer simulations to investigate 
the ionization of this system. We also
analyze the relative metal abundances and comment 
on the characteristics of dust
grains in this system.  In addition, we examine a \Lya absorption system
at $z$ = 1.955 and use kinematic and 
abundance arguments to suggest it may be a
damped \Lya system.  Finally, we discuss the kinematics 
of the $z$=2.462 damped
absorption system, contrasting its features with several 
other systems toward
Q0201+365, as well as other damped \Lya systems measured with HIRES 
(\cite{wol94}).

\begin{figure}
\centerline{
\psfig{figure=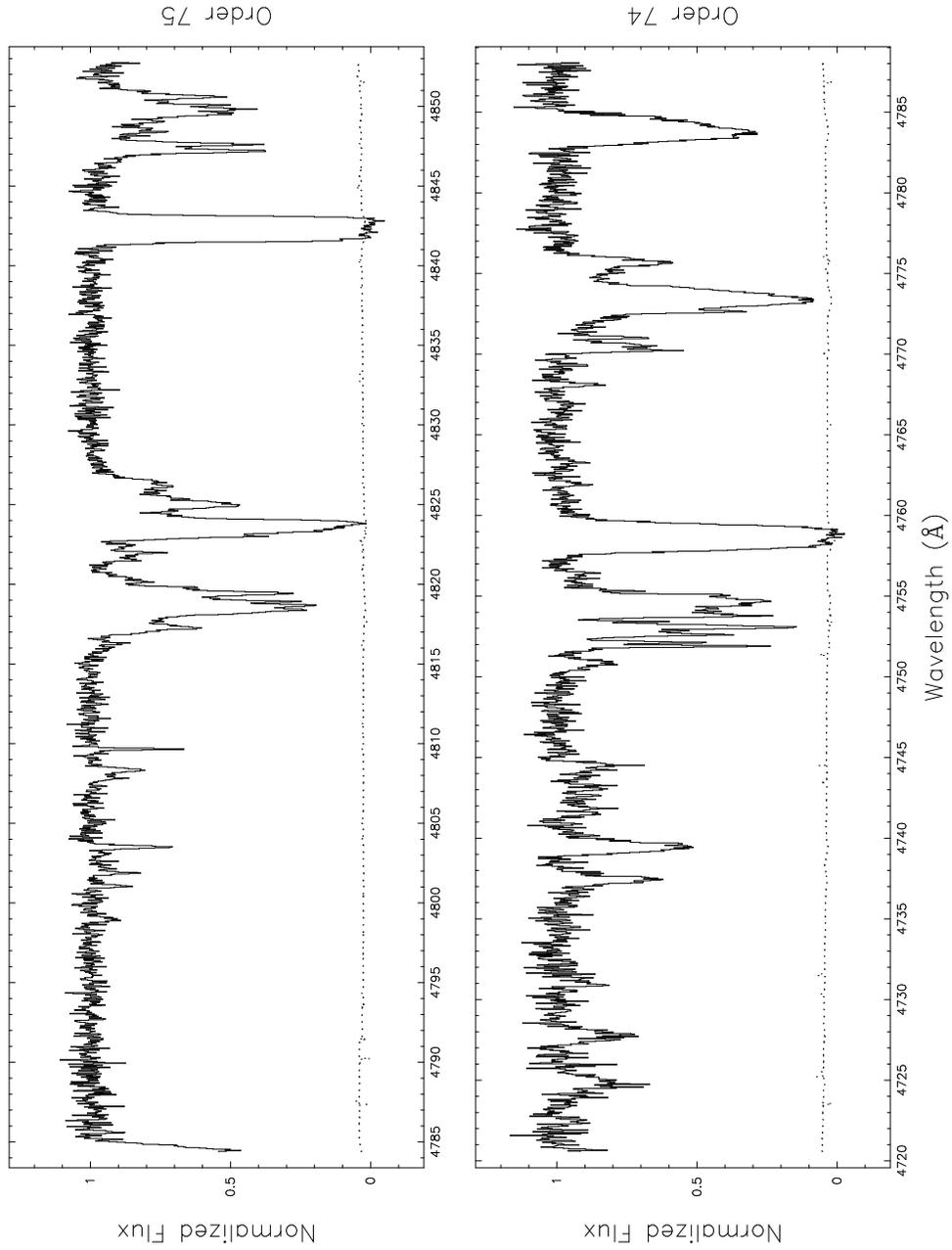,height=7.5in}}
\caption{ Keck HIRES spectra of Q0201+365 at a resolution of $\approx 8$ \kms
and a SNR of $\approx 30$.  All 26 orders (only 2 are shown here)
are identified by the
echelle order intrinsic to the Keck HIRES spectrograph.  The dotted line is a
1$\sigma$ error array derived assuming Poisson statistics.}
\label{sptra}
\end{figure}

     The paper is organized as follows.  In $\S$ 2 
we describe the data acquisition
and reduction techniques, present the spectra 
(Figure~\ref{sptra}) and give a nearly
complete absorption line list in Table~\ref{orders}.  
We detail the analytic methods utilized
throughout the paper in $\S$ 3.   In $\S$ 4 we present 
velocity profiles of the most
significant metal line systems along with the VPFIT package solutions where
applicable.  $\S$ 5 presents the ionic column densities of the two systems
associated with the damped \Lya profile at $z$=2.46.   
In $\S$ 6 we argue that the
degree of photoionization of the damped \Lya system is low.  
$\S$ 7 gives the results of the abundance measurements and 
discusses the possible depletion of the
gas-phase metals by dust grains.  We also describe the kinematics 
of the \Lya
system. Finally, $\S$ 8 summarizes the results 
and gives concluding remarks.

\section{DATA}

     In this section we present the HIRES spectra, detailing the
techniques used for the acquisition and reduction of the data. 
Table~\ref{orders} gives an absorption line list with measured equivalent
widths and 1$\sigma$ errors and identifies over 80$\%$ ($\geq 88\%$ redward
of the \Lya forest) of the features.

\subsection{Acquisition}

     We observed Q0201+365 with the HIRES echelle spectrograph
on the 10m W.M. Keck Telescope on three separate nights for a 
total integration
time of 9.7 hrs.  Table~\ref{obs} presents a journal of the 
observation dates, exposure
times, wavelength coverages and resolution of the data.  
Unfortunately, the signal
to noise ratio (SNR) of the data was significantly limited 
by clouds.  We used the
kv38 filter to block out 2nd order blue wavelengths, the C5 
decker plate with
1.1$''$ slit, and standard 2$\times 1$ binning on the 
2048 x 2048 Tektronix CCD. 
This setup afforded a resolution ranging from $7.2 - 8.0$ \kms 
and wavelength coverage from $4720-7180 \rAA$ over 26 orders. 
Gaps are evident between
orders redward of $\approx 5100 \rAA$ where the free 
spectral range of the
echelle exceeds the format of the CCD.  For reduction and calibration 
we took
300s exposures of the nearby standard star BD+28411 and 
images of quartz and Th-Ar arc lamps. 

\subsection{Reduction}

     The 2-D CCD images were reduced to 1-D uncalibrated spectra
with a software package kindly provided by T. Barlow (1995).  In
short, the package subtracts the baseline and bias from the 2-D
frames, determines the gain from dark images converting digital
numbers (DN) to e$^-$ counts, and extracts the data from the 2-D
images by tracing the bright standard-star profile.  It also performs sky
subtraction and the removal of cosmic ray events. 

     We wavelength calibrated the data of each night by
fitting low-order Legendre polynomials to our Th-Ar arc frames,
properly correcting to vacuum heliocentric wavelengths.  
We obtained continuum 
fits to the spectra by fitting high-order Legendre 
polynomials to the spectral 
flux in each order.  The continuum fits were then used to 
normalize the flux 
to unity.  We also calculated a 1$\sigma$ error array assuming Poisson 
statistics, ignoring the errors associated with sky subtraction 
and continuum fitting (i.e. $\sigma =
\sqrt{N_{\rm obj} + N_{\rm sky}}$).  Finally we coadded the 
spectra, weighting
by the calculated SNR of each image while rejecting spurious 
values assumed to
arise from cosmic rays or poor sky subtraction.  The final average SNR is
$\approx$ 30 with the exception of Order 75 (SNR $\approx 22$) 
where only one
night of data was acquired due to a difference in the alignment of the
CCD on the first night.  The spectra are presented along with
1$\sigma$ error arrays in Figure~\ref{sptra}. 

\subsection{Absorption Lines}

     Table~\ref{orders} lists the wavelengths, equivalent 
widths and 1$\sigma$ errors
for all absorption line features which exceed the 
5$\sigma$ limit in equivalent
width as measured by techniques similar to those of Lanzetta 
et al. (1991).  We
believe 5$\sigma$ is a conservative but appropriate limit for this data. The
reported wavelengths represent rough estimates of the 
centroids of complex line
profiles and should be taken only as guides 
for differentiating between features. 
Because almost all of the transitions are resolved, 
an accurate determination of 
the wavelength of every feature would require months of 
laborious Voigt profile
fitting.  This task bears little scientific merit for the 
task of the present paper
which is to examine specific systems.  For this reason, 
we carried out detailed
profile fits for only a selected subset of 
the lines in Table~\ref{orders}.  We do not report
equivalent widths for those features identified as sky absorption lines or
complicated multi-system blends. 

     Table~\ref{orders} also includes the transition names and approximate
redshifts of those features we successfully identified.  
Identification proceeds in a
largely {\it ad hoc} fashion, with the initial emphasis placed on
finding C IV and Si IV doublets.  Once we composed a list of redshifts for the
metal-line systems, we attempted to match the remaining features with the
strongest metal-line transitions.  Finally, we compared 
the line profiles of the
transitions of a given redshift system in velocity space for conclusive
identification.  Although fairly laborious, this approach 
is highly effective and essentially error-proof.

     By comparing the object frames with the similarly reduced
standard star images, we identified night-sky emission and
absorption features in the spectra.  These features are labeled
appropriately in Table~\ref{orders} 
along with all other identified spurious features.  

	Table~\ref{fval} lists the rest wavelengths and oscillator
strengths for all of the metal transitions analyzed in this paper.  Almost
all of the values are taken from Morton (1991).  It is
important to be very clear what values are assumed in order to 
allow for consistent abundance comparisons.

\section{ANALYTIC METHODS}

     This section describes the least squares line profile
fitting method and the apparent optical depth method used to
analyze our metal line systems.

\subsection{VPFIT (Least-Squares Line-Profile Fitting Method)}

     With the aid of the VPFIT fitting package
kindly provided by R.F. Carswell we performed
least-squares fits of Voigt profiles, produced by 
individual Gaussian components,
to our metal lines for several of the absorption systems 
toward Q0201+365.  The
VPFIT package fits Voigt profiles to an absorption system, simultaneously
determining the redshift, column density, $b$ values 
(where the Doppler velocity $b$
and velocity dispersion $\sigma$ are related by $b = \sqrt{2} \sigma$) and
associated errors of the individual components while minimizing a $\chi^2$
parameter matrix .  When performing a fit, one can tie several transitions
together forcing the redshift and $b$ values of the different 
transitions to match
while allowing the column densities to vary individually.

     In performing the fits we first isolated the broadest 
resolved transition in a
given absorption complex.  We found we could always obtain a reasonably
accurate profile fit to this transition.  We then tied it together 
with the other
associated transitions.  Given the inherent differences between low and 
high-ion line profiles, we chose to treat the two 
types separately.  If necessary, we
added or removed a velocity component to those transitions with line profiles
that have features not evident in the other transitions.  In several 
cases, this
significantly lowered the final $\chi^2$ value.  In all our fits 
we assumed bulk
motion dominates thermal motion because damped \Lya systems 
are relatively
cool ($T <$ 1000 K) and the transitions arise in metals 
that are comparatively
massive.  We also set a minimum value for the $b$ parameter 
at 3 \kms to
prevent the package from fitting features narrower than 
$\approx$ 4 pixels, the
FWHM of the line spread function.

\subsection {Apparent Optical Depth Method and Hidden Component
Analysis}

     Savage and Sembach (1991) have stressed that measuring
column densities with the line profile method does not always account for
hidden saturated components.  These saturated components may be
under represented in the line profile analysis, leading to
significant errors in the measured ionic column densities.  
Therefore, as a check
on our VPFIT results we performed a hidden component analysis of those ions
where multiple transitions were observed (e.g. Ni$^+$ for the damped \Lya
system at $z$ = 2.462). The analysis involves calculating $N_a (v)$, 
the apparent
column density per unit velocity, for each pixel from the optical 
depth equation

\begin{equation}
N_a(v) = {m_e c \over \pi e^2} {\tau_a(v) \over f \lambda} , 
\end{equation}

\noindent where $\tau_a(v) = \ln [I_i (v) / I_a (v)]$, f is
the oscillator strength, $\lambda$ is the rest wavelength and
$I_i$ and $I_a$ are the incident and measured intensity.  Comparing $N_a
(v)$ deduced from two or more transitions of the same ion, one 
finds the stronger
transition will have smaller values of $N_a (v)$ in those features where
hidden
saturation is present. Thus, one can ascertain the likelihood of saturated
components for ions with multiple transitions.  Furthermore, summing $N_a(v)$
over the proper velocity intervals serves as an excellent 
check on ionic column
densities measured with the VPFIT package, and as a further check on the
existence of saturation.

\section{VELOCITY PLOTS AND VPFIT SOLUTIONS}

     This section presents the velocity profiles
of the most complex metal line absorption systems toward
Q0201+365.  For several of the systems, we superimpose the
solutions of our least square fits from the VPFIT package.  In 
all plots a dashed
vertical line is drawn for reference,  usually identifying 
the strongest feature
present.  For clarity, we have plotted features not related to 
the given system
with dotted lines.

\subsection{Damped \Lya System ($z$ = 2.462)}

\begin{figure}
\centerline{
\psfig{figure=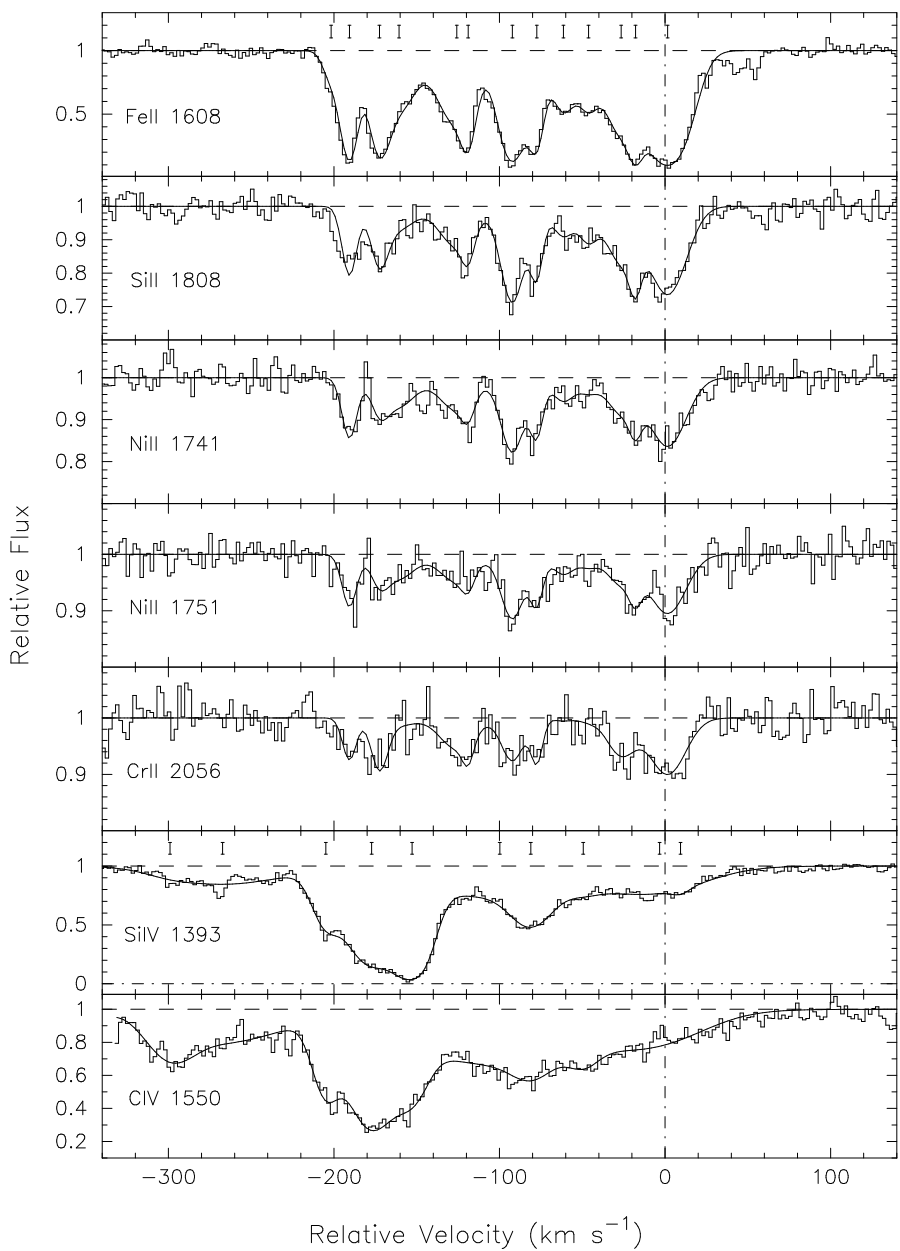,height=7.5in}}
\caption{ Velocity profiles of those metal-line transitions 
from the system at $z$=2.462 with successful VPFIT solutions.
The dashed vertical line corresponds to $z$=2.4628.  
The marks
above the Fe II 1608 and Si IV 1393 transitions indicate the 
velocity centroids for 
VPFIT solutions as listed in Table 4.  
In all velocity plot figures with VPFIT solutions,
the leftmost mark is component 1.}
\label{2462V}
\end{figure}

     Figure~\ref{2462V} presents the line profiles 
and fits of the transitions associated with
the damped \Lya system at $z$=2.462\footnote{ Note 
there are two `subsystems' within
the velocity space of the damped \Lya profile: the present subsystem 
and the $z$=2.457 subsystem discussed in $\S$4.2}.  
The velocity centroids for the Gaussian
components of the fit to the low-ion profiles are denoted 
by the short vertical lines
above the Fe II 1608 profile.  The velocity centroids for 
the high-ion fits are the
vertical lines above the Si IV 1393 profile. We found 13 individual
components were required for an optimal fit to the low-ions.  The
multi-component fit to the transitions Fe II 1608, Ni II 1741, 
Ni II 1751, Cr II
2056 and Si II 1808 has a reduced $\chi_\nu^2 = 1.05$ 
with a probability $P_{\chi^2} = 0.216$.  Table~\ref{2462TL} 
lists the redshift, $b$ value and column density along with
1$\sigma$ error of every velocity component in the VPFIT solution.  
The quality of the fit reflects the high degree 
to which the low-ion profiles track one another.  

     On the other hand, 10 components (only 8 over the same 
region as the low ion
profiles) were necessary for an optimal fit to the
high-ion transitions C IV 1550 and Si IV 1393. 
These components are significantly
broader (higher $b$ values) than those of the low-ions.  This 
fit is not as accurate as the low-ion fit ($\chi^2_\nu = 1.805$)
because it is much more difficult to fit 
broad shallow components such as those around $\approx -260$ \kms. The
results of this fit are presented in Table~\ref{2462H}. 

\begin{figure}
\centerline{
\psfig{figure=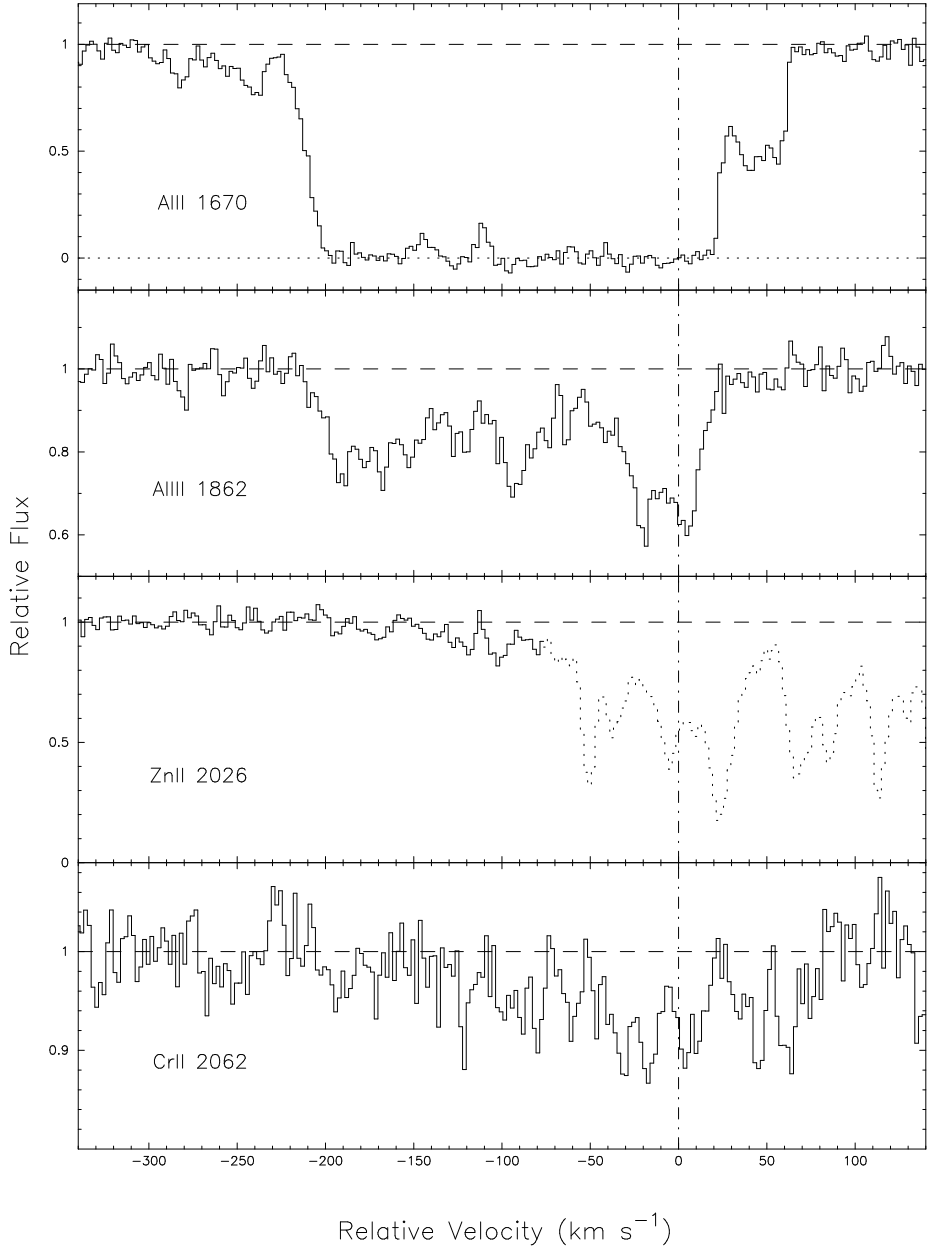,height=7.5in}}
\caption{ Velocity profiles of those metal-line transitions from the
system at $z=2.462$ without VPFIT solutions.  The dashed vertical
line corresponds to $z$=2.4628.}
\label{2462Vb}
\end{figure}

     Figure~\ref{2462Vb} shows velocity profiles of 
4 transitions we chose not to fit with the VPFIT package.  
Because the Al II 1670 profile is so highly
saturated, we were unable to accurately fit it with the 
other low-ion profiles.  In particular, the VPFIT package 
could not properly model the heavy absorption at
$\approx -140$ \kms. The Al III 1862 profile 
exhibits characteristics of both the
low and high-ion transitions and therefore was difficult 
to fit it to either.  Instead, we measured its column 
density with the apparent optical depth method.  Finally,
the 3 transitions Zn II 2026, Cr II 2062 and Zn II 2062 
all suffer from significant blending: Zn II 2026 is 
blended with both Fe II 2374 and Fe II 2600 from 2 other
systems, and Zn II 2062 and Cr II 2062 are blended 
with each other.  As a result
all three transitions were excluded from the fit.

\subsection{Companion System at $z$=2.457}

\begin{figure}
\centerline{
\psfig{figure=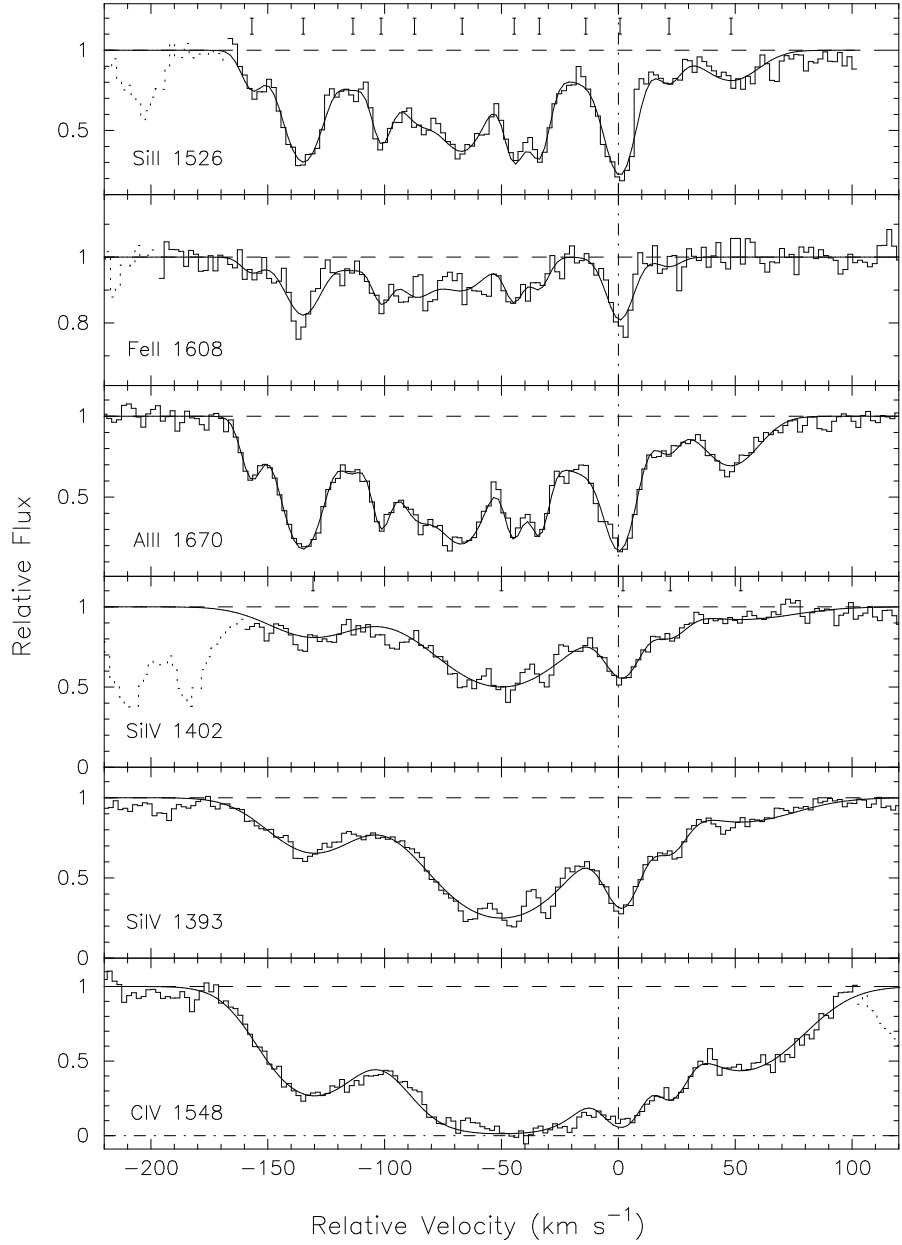,height=7.5in}}
\caption{ Velocity profiles and VPFIT solutions for the transitions 
from the system
$z$=2.457.  The dashed vertical line corresponds to $z$=2.45785.  
Those features drawn in 
dotted marks are due to absorption from other systems.}
\label{2457V}
\end{figure}

     The velocity profiles and least-square fits of the other
subsystem (found at $z$=2.457) associated with the damped \Lya
system at $z$=2.46 is shown in Figure~\ref{2457V}.  A 12 component fit to
the low-ion transitions Fe II 1608, Si II 1526 and Al II 1670 
was optimal and yielded a reduced $\chi_\nu^2 = 1.50$. 
Difficulties arose in fitting the Fe II
1608 profile because the second component appears 
to have a significantly lower
$b$ value than the same component in the other low-ion profiles.
This may be a result of thermal broadening. We
also fitted the high-ions SiIV 1393, 1402 and CIV 1548 
and again found broader,
fewer components were required in the best fit.  Tables~\ref{2457L} 
and \ref{2457H} list the results of the fits.

\subsection{Possible Damped System at $z$=1.955}

\begin{figure}
\centerline{
\psfig{figure=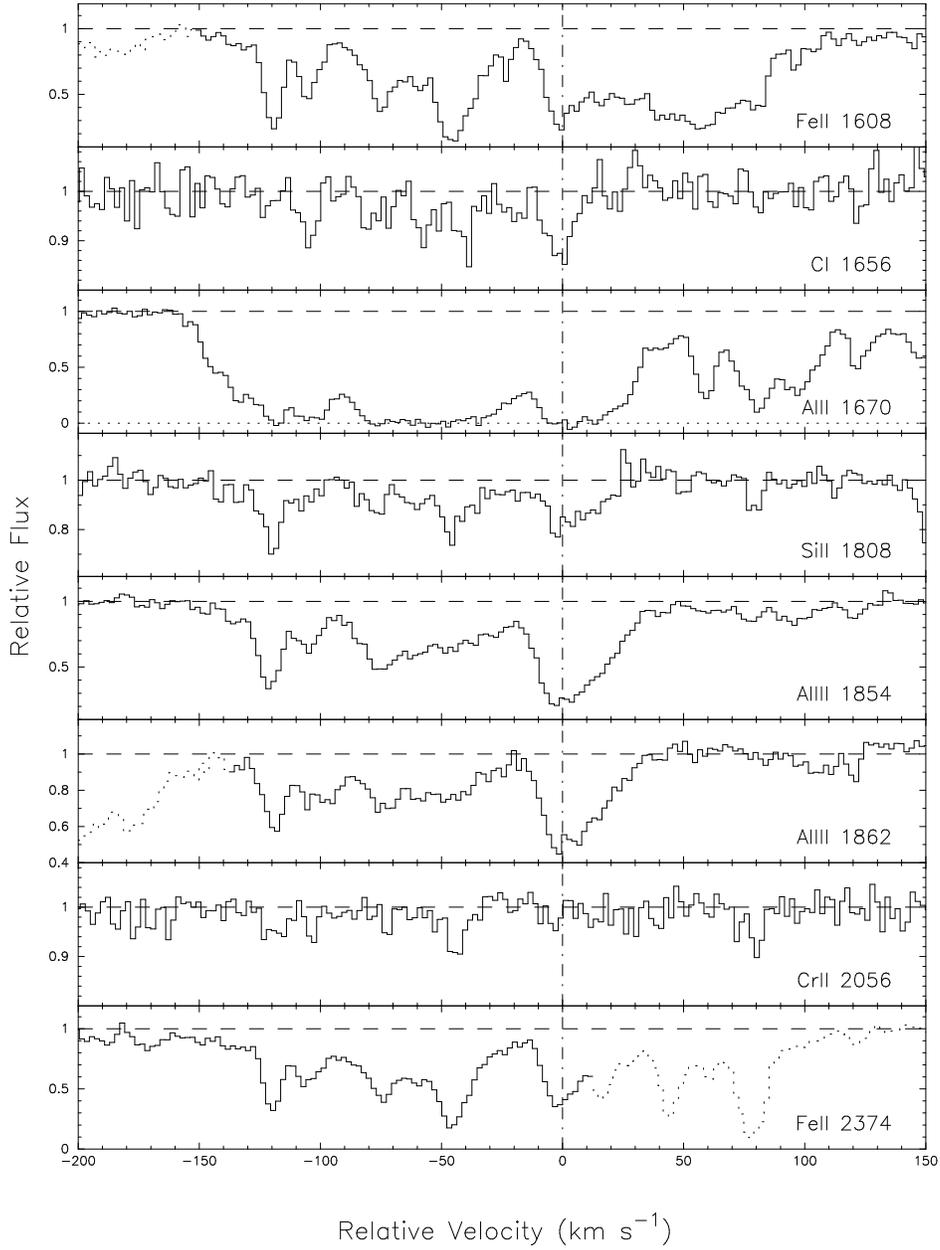,height=7.5in}}
\caption{ Velocity profiles for the metal-line 
transitions for the system at $z$=1.955.  The
dashed vertical line corresponds to $z$=1.9555.}
\label{1955V}
\end{figure}

     Figure~\ref{1955V} presents the velocity 
profiles for a redshift system with 
$z$=1.955 which is a possible damped \Lya system.  We chose
not to fit this system because significant blending with other systems
prevented us from determining which features are associated only with
the $z$=1.955 system.  In particular, we have been unable to determine
whether the absorption seen at positive relative velocities 
in Figure~\ref{1955V} is due to the $z$=1.955 system.

\begin{figure}
\centerline{
\psfig{figure=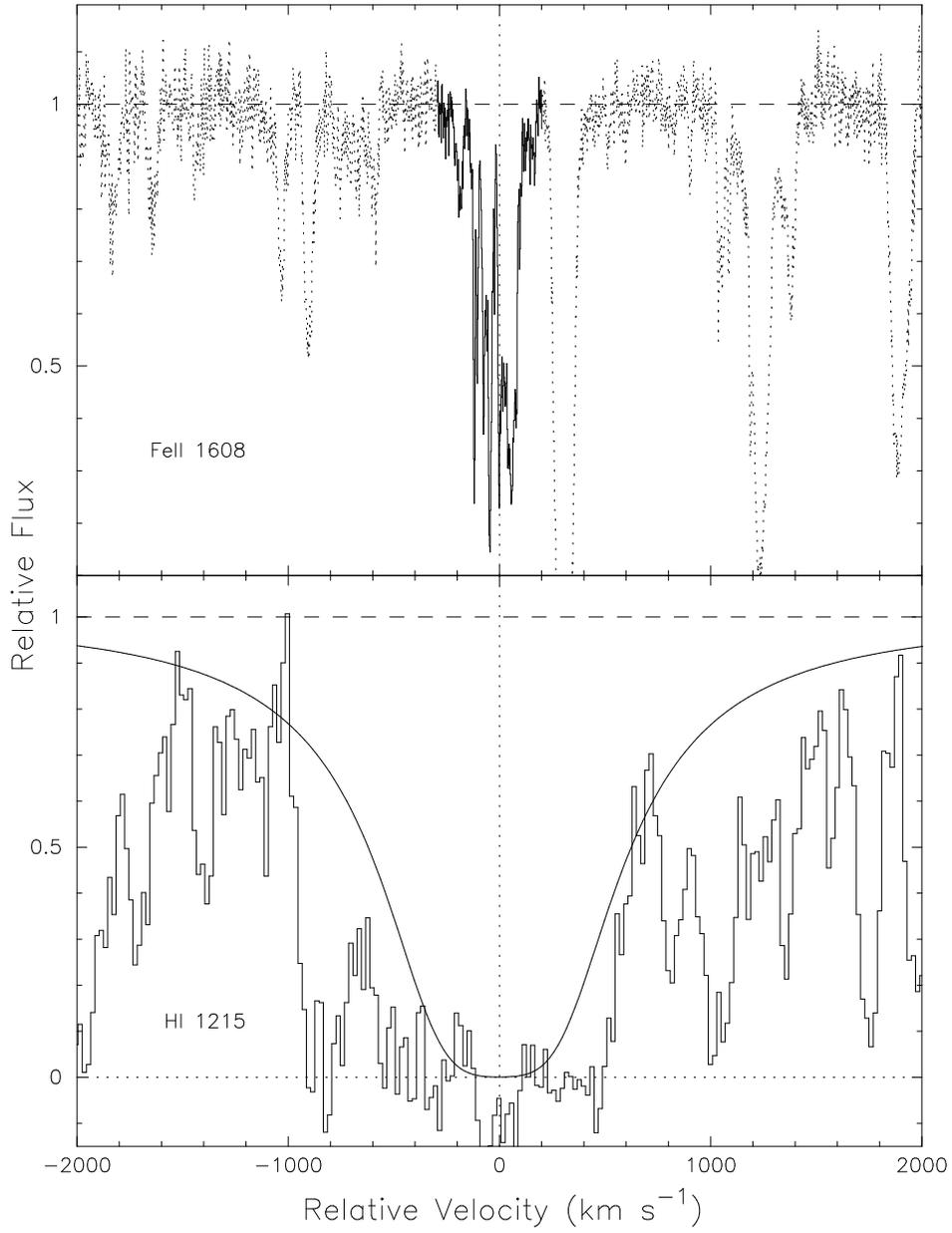,height=7.5in}}
\caption{ Comparison of the low-ion profile Fe II 1608 with the 
\Lya profile for the
system at $z$=1.955.  The overplotted curve in the \Lya plot is a 
Voigt profile
corresponding to $z$=1.9555 and $\N{HI} = 1.5 \sci{20} \cm2$.  }
\label{1955L}
\end{figure}

	Lu et al. (1994) could 
not confidently classify the system as damped because the noisy \Lya 
profile could not be fitted with a Voigt damping profile.  The metal 
lines have many resolved components and exhibit a large velocity 
interval ($\approx$ 250 \kms).  Figure~\ref{1955L} plots the observed \Lya 
profile, a Voigt damped \Lya profile, and the
Fe II 1608 line profile.  The Voigt \Lya profile was derived by letting 
$\N{HI} = 1.5 \sci{20} \cm2$ at $z$=1.955.  It appears the low-ions 
satisfy the metal criteria associated with damped \Lya systems; i.e., 
the low-ion metal profile is significantly narrower than the \Lya
profile and its velocity centroid is near that of the \Lya profile 
(Wolfe et al. 1993).  An apparent optical depth measurement of the 
Cr II and Si II ions gives $\N{Cr^+} = 12.73 \pm .070$ and $\N{Si^+} = 15.25
\pm 0.03$ suggesting a lower limit of $\log[\N{H}] \geq 19.0$ and $19.7$
respectively, assuming cosmic abundances and no dust depletion.  
This analysis
indicates the $z$=1.955 system may be damped.  Furthermore, 
Figure~\ref{1955V} reveals evidence for C I 1656, 
a transition previously seen only
in damped \Lya systems.  However, even with our high
resolution data the proper classification remains inconclusive.  

\subsection{System at $z$=2.325}

\begin{figure}
\centerline{
\psfig{figure=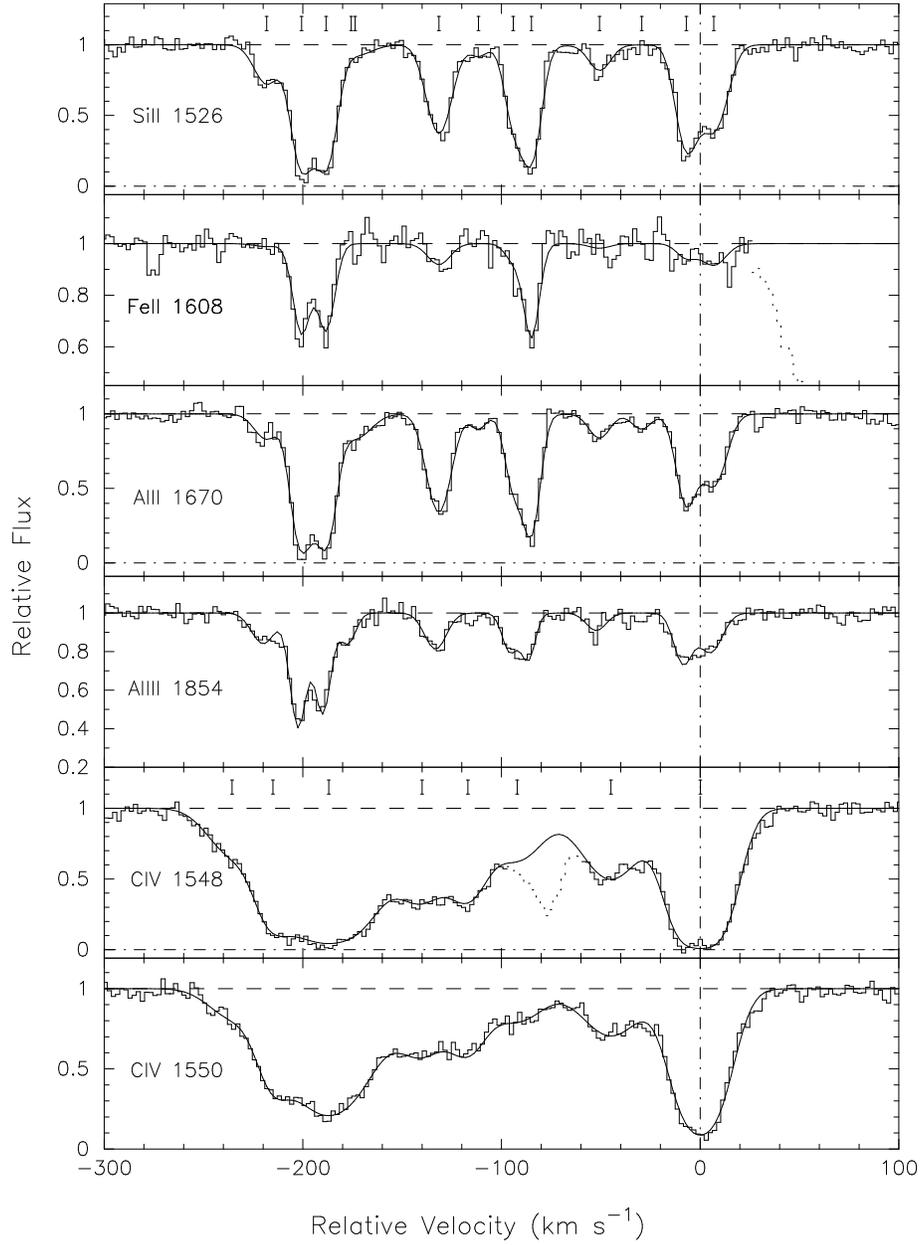,height=7.5in}}
\caption{ Velocity profiles and VPFIT solutions from the system 
at $z$=2.325.  The
dashed vertical line corresponds to $z$=2.32612.  }
\label{2325V}
\end{figure}

     The velocity profiles and VPFIT profile solutions for
the absorption system at $z$=2.325 are shown in Figure~\ref{2325V}.  
Note the feature at $-80$ \kms in the CIV 1548 transition is due to a
blend with the CIV 1550 transition from the system at
$z$=2.320.  By contrast with the damped system, 
the Al III 1854 transition was best fitted with the 
low-ion solution.  We tied the Si II 1526, Fe II 1608, Al II 1670,
and Al III 1854 transitions together and fitted the CIV doublet separately.  
Tables~\ref{2325TL}  and \ref{2325H} 
present the results of the 2 fits.  Unlike the damped system, the high and
intermediate ions more closely follow the low-ion profiles. 
Because the high-ion profiles have several 
components with higher $b$ values, however, it is still
impossible to fit them together.  

\begin{figure}
\centerline{
\psfig{figure=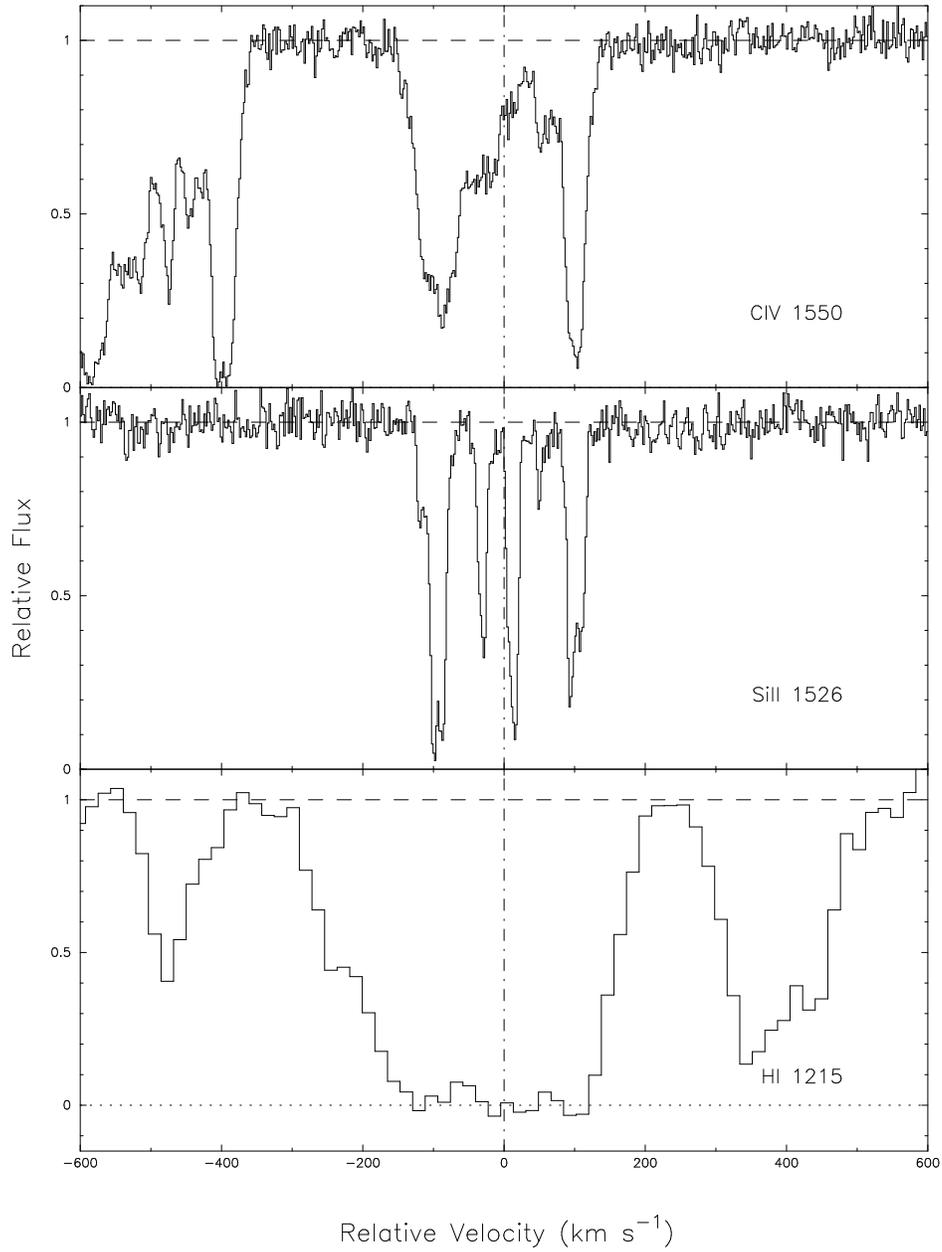,height=7.5in}}
\caption{ Velocity plot of the C IV 1550, Si II 1527 and \Lya transitions
for the system at $z$=2.325.  The transitions all span nearly the same velocity
interval and appear consistent with profiles of Lyman limit systems.}
\label{2325L}
\end{figure}

	Figure~\ref{2325L} plots the Si II 1526, C IV 1550 
and \Lya profiles for the $z$=2.325 system.  All three transitions
span nearly the same velocity interval and track one another moderately
well.  As expected the \Lya profile tracks
the low-ion profiles (in particular the component at $v \approx 20$ \kms) 
more closely than the high-ion profiles.  
Although we did not fit the \Lya profile, Figure~\ref{2325L}
is consistent with the profiles of Lyman limit systems. 
     
\subsection{Mg II Systems at $z$=1.476, 1.699}

\begin{figure}
\centerline{
\psfig{figure=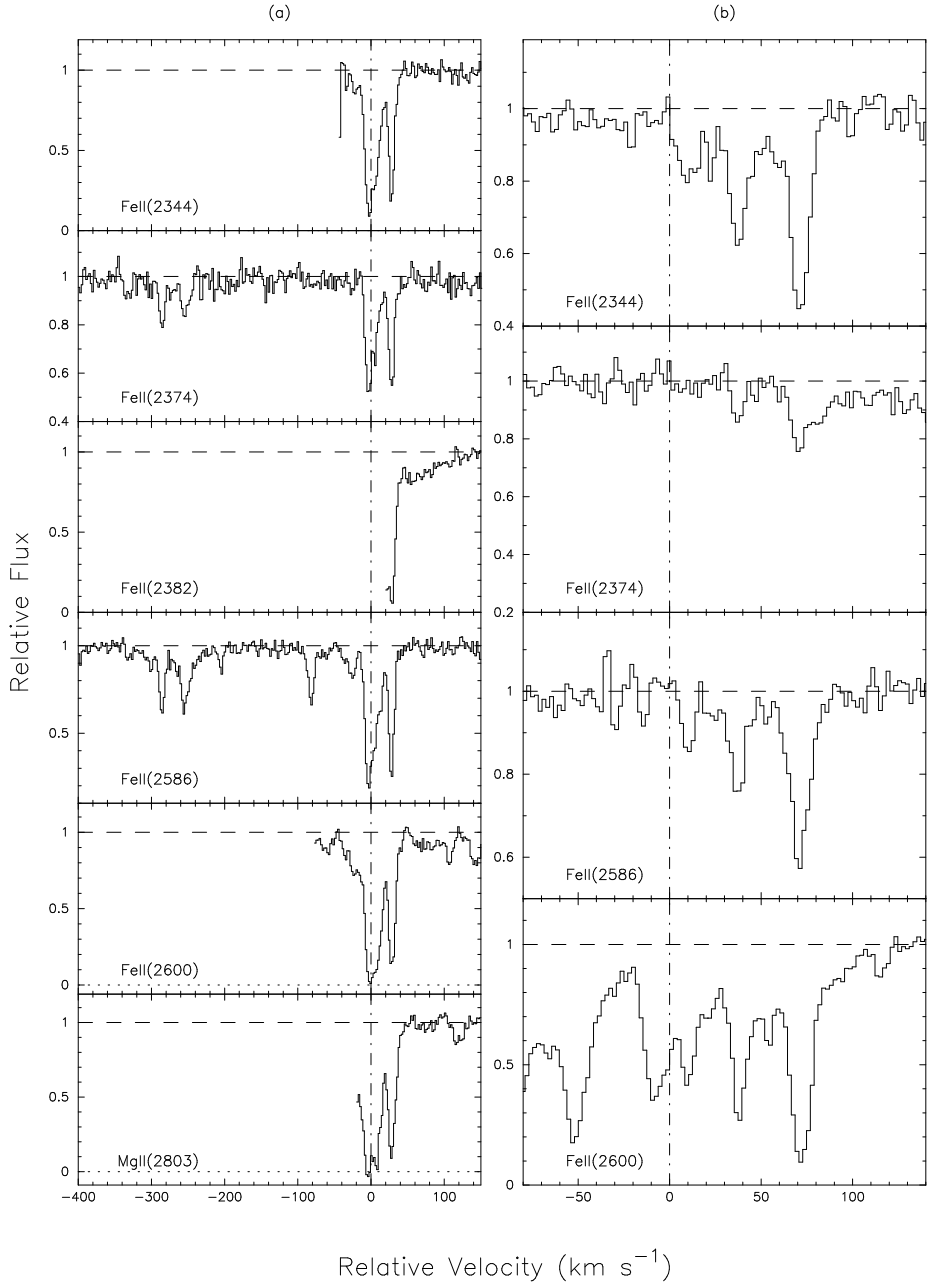,height=7.5in}}
\caption{ Velocity profiles of two Mg II metal systems at 
(a) $z$=1.476 and (b) $z$=1.699.  The dashed
vertical lines are at $z$=1.47607 and $z$=1.699 respectively.}
\label{MgII}
\end{figure}

     Figure~\ref{MgII} shows the velocity profiles of two Fe II 
and Mg II transitions for metal line systems found at the 
redshifts $z$=1.476
(a) and $z$=1.699 (b).  Both sets of transitions are relatively complex and 
most of the absorption spans moderate velocity intervals 
($\approx 100$ \kms).  However, the Fe II 2374 and Fe II 2586
transitions in the $z = 1.476$ system may exhibit an additional feature at
$\approx -275$ \kms.  
Because the predicted wavelengths for the other Fe II lines
fall in the inter order gaps, we cannot verify the reality of this feature.

     Figures 10a and 10b are hidden component analyses of the
Fe transitions from the two systems.   Fe II 2586 was 
not plotted in either hidden component analysis
because it closely traces the Fe II 2374 profile in each system and
would clutter the figures. In Figure 10a, 
the weakest Fe transition (Fe II 2374) 
has significantly larger $N_a (v)$
values, while the strongest 
Fe transition (Fe II 2600) has the smallest $N_a (v)$
profile.  This is direct evidence of hidden saturated components.  In fact, 
because of these hidden components, we found it impossible to fit all of the
transitions together in the $z$=1.476 system.  
We were successful in fitting the Fe
II 2344, 2586 transitions, however, 
and the results are presented in Table~\ref{1476}.  

     There is little evidence of hidden saturated components in 
the $z$=1.699 system. What the hidden component analysis does reveal, 
however, is a blend in the Fe II 2374 profile with Al III 1854
from the $z = 2.457$ at velocities greater 
than $\approx$ 0 \kms and a blend in the Fe II 2600 profile with Fe II 2374
from the $z$ = 1.955 system at velocities 
lower than $\approx$ $-70$ \kms.  Unfortunately, these blends, and the
low SNR in the Fe II 2586 and Fe II 2374 profiles prevented accurate
profile fits.

\section{IONIC COLUMN DENSITIES}

     This section presents the ionic column densities of the 
two systems associated
with the damped \Lya profile at $z=2.46$. We perform hidden component
analyses where applicable, and use both the line profile and 
apparent optical
depth methods to measure column densities.

\subsection{Damped \Lya System at $z$ = 2.462}

     The hidden component analysis of the Ni II 1741, 1751 
transitions for the
damped \Lya system at $z$ = 2.462 is shown in Figure~\ref{HCA-Ni}. 
With a few minor exceptions, the $N_a (v)$ curves 
for the two Ni II transitions match within $1
\sigma$, suggesting no hidden saturated components.  Therefore, abundances
based on column densities inferred from profile fitting 
or the apparent optical depth method should not suffer 
from significant hidden
saturation effects.  Table~\ref{2462I} lists the measured
ionic column densities and 1$\sigma$ errors for 
Fe$^+$, Cr$^+$, Si$^+$, Al$^+$ and Ni$^+$ for the 
damped \Lya system at $z$=2.462 as measured by both the
line profile (VPFIT) and apparent optical depth methods.  
For the line profile
method we summed the column densities of the 
individual components of each transition and calculated 
the 1$\sigma$ error in the total value with standard
least squares techniques.  

     We adopt a final value for the ionic column density for all of our
measurements by averaging the two values and adopting the VPFIT errors.  We
found the apparent optical depth method underestimates the 
error, particularly
transitions which are nearly saturated (e.g. FeII 1608).  
For nearly saturated or very weak lines, 
errors associated with sky subtraction and continuum fitting will play a
significant role, yet these errors are not included 
in the 1$\sigma$ array.  The VPFIT package estimates 
errors based on the fit of all profiles and is influenced
more by the deviation of the data from the fit.  
Therefore, we chose to adopt the VPFIT errors for all of the ions.  Note 
that in almost every case the calculated values
from the two methods match, further 
indicating no hidden saturated components.
 
     As noted above, we can place only a lower limit on the
Zn$^+$ column density because of blending. 
We find the Zn II 2026 transition to be dominated by 
Fe II 2374 associated with the $z$ = 1.955 absorption system, and
Zn II 2062 to be partially blended with the stronger Cr II 2062 transition.
Although Zn II 2026 cannot be extracted from 
the Fe II profile, the Zn II 2062
profile can be used to estimate a lower limit on the Zn$^+$ abundance.  
Figure~\ref{HCA-Cr} is a hidden component 
analysis of the Cr II 2056 and Cr II 2062 transitions.  
Note that the Cr II 2062 (dotted) profile dominates the 
Cr II 2056 profile over the entire velocity interval, as 
expected if Zn II 2062 is present.  This analysis reveals
two features around 40 and 60 \kms evident only in the 
Cr II 2062 profile.  We suggest that these two 
components are not due to Cr$^+$ absorption, but are
components of Zn II 2062 at the high velocity edge of the profile.    
In the velocity space of Zn II 2062 these components 
correspond exactly to the strongest components of the 
other low-ion profiles at $z$=2.46258 and $z$=2.46280 
(i.e. at v = $-20$ and 0 \kms in the Zn II velocity space).  
Our hidden component analysis of Cr II reveals a third 
component of Zn II at 7139.66 $\rAA$ 
($-60$ \kms in Figure~\ref{HCA-Cr}), 
also corresponding to a significant feature in the other low-ion profiles.

     Summing the column densities over velocity space with the 
apparent optical depth method, we find
we find $\log \N{Zn^+}_i = 11.753 \pm 0.116, 12.087
\pm 0.066,$ and $12.060 \pm 0.074$ respectively for the three components
(blue to red).  
The resulting lower limit for the total
Zn$^+$ abundance, $\log \N{Zn^+}_t = 12.468 \allowbreak \pm 0.046$ where the
error reported is only useful as an indication of the uncertainty 
in this lower limit value.  

       Table~\ref{2462I} also lists ionic column densities 
for the $s$-process transitions
Pb II 1682 and Ge II 1602.  
Because these transitions are so weak, we the used linear
curve of growth to infer the column densities from the respective 
rest-frame equivalent widths.   
We report the following values for the rest-frame 
equivalent width of the transitions:
$W({\rm Pb}) = 1.78 \pm 0.47 \sci{-2} \rAA$ and $W({\rm Ge}) = 
-3.4 \pm 4.6 \sci{-3} \rAA$.  
The negative equivalent width value for Ge II 1602 
is certainly a function of the
continuum error and is consistent with a null
detection (less than even a $1 \sigma$ detection).  We have chosen, therefore,
to report its column density and abundance in terms of 
a $3 \sigma$ upper limit.
The Pb II 1682 value is significant at the $3 \sigma$ level, 
but could also be explained through a sizable continuum error, 
improper sky subtraction, or an unidentified blend. 
Although Sn II 1400 and Ga II 1414 both lie within our wavelength coverage,
they are overwhelmed by blends with transitions from other 
metal line systems and could not be analyzed.

\subsection{Companion Subsystem at $z$=2.457}

     Table~\ref{2457I} presents the measured ionic 
column densities for the transitions of the
companion system at $z$=2.457.  Final values are adopted according to the
criteria described above.  
In nearly all of the low-ion transitions the measured
column densities are $\approx$ 20 times lower than those from the $z$=2.462
subsystem.  On the other hand the column densities of the high-ions are nearly
the same, possibly indicating this system is in 
a higher state of ionization, that
the two ion groups are kinematically disjoint, or that the same high-ion gas
envelopes two dissimilar low-ion configurations.

\section{IONIZATION}

     This section investigates the 
photoionization of the damped \Lya system at $z$=2.462.

\subsection{\Lya Profile}

     Figure~\ref{2462L} is a velocity plot of the 
\Lya profile and Fe II 1608 profiles
associated with the two absorption systems 
at $z$=2.462 and $z$=2.457 together with
the velocity profile of the Fe II 1608 transitions.  
Unrelated features at $\approx -800$ and 500 \kms have been 
removed for clarity.  We derived the Voigt profile
by distributing the total HI column density, 
$\log \N{H^0}$ = 20.38, into the 23 Fe II 1608 velocity 
components weighted by their corresponding column density fractions
($\N{Fe^+}_i / \N{Fe^+}_{\rm tot}$).  This treatment is proper
provided (a) the Fe II 1608 line profiles accurately trace 
the HI gas, as one would
predict for a sufficiently neutral system, 
and (b) [Fe/H] is constant across the
entire velocity profile, which is 
predicted for a well mixed system.  As Figure~\ref{2462L}
demonstrates, the resulting Voigt profile is 
well fitted to the low resolution data.
Table~\ref{HI} lists the column density, 
redshift and $b$ values of the 23 components as
adopted.  Here, about 5$\%$ of the HI gas 
is located in the companion system at
$z$=2.457.  We found one could place no more than 15$\%$ of the HI 
gas in the
companion system before significantly distorting the left wing 
of the \Lya
line profile.

\subsection{Ionization Models}

\subsubsection{Neutral Hydrogen Model}

     Table~\ref{HI} shows that all but the 
most abundant components of our damped
system have $\log[\N{HI}] \leq 19.5$.  
If these clouds were isolated structures in the
IGM, they would be highly ionized.  This would markedly reduce the
accuracy of abundance determinations based on the 
assumption that most of the hydrogen
is neutral and most of the metals are singly ionized.

     In order to address this problem we ran several 
ionization simulations with the aid of
a program developed by Vincent Virgilio to investigate the predicted neutral
hydrogen fraction (H$^0$ / H) in the damped system.  Figure 14a shows the
neutral hydrogen fraction plotted against 
$\log[\N{H^0}]$ measured from either face of
a plane-parallel layer with constant H volume density, $n_H = 0.1 \> \rm
cm^{-3}$.  The layer is subjected on both sides to attenuated power
law continuum radiation as calculated by 
Madau (1992) for a redshift of $z$=2.46 with a
mean intensity of 
$J_\nu = .195 \sci{-21} \> \rm ergs \> s^{-1} \cm2 \> Hz^{-1}
\> sr^{-1}$ at 1 Rydberg.  The simulation assumes a temperature of $10^4$ K
(purely for determining the recombination rate of H$^+$)
while satisfying the ionization and 
transfer equations in a large number (100) of
parallel discrete cells.  Each face of the 
plane parallel layer is illuminated by
uniform radiation with incidence angles covering $2 \pi$ steradians.  The
calculation also assumes zero source function (i.e. no reemission).
Although this model is rather simplified, we find
similar CLOUDY (version 84.12: Ferland 1991) calculations are in good
agreement with our results.  Figure~\ref{I-Comp} compares the neutral
fraction predictions of
our model with the corresponding predictions by CLOUDY.
The 2-sided illumination simulation
predicts a lower neutral fraction than CLOUDY even though
CLOUDY assumes only perpendicular incidence.  
Thus, our simulations reveal that the 1-sided assumptions
inherent in CLOUDY are probably
underestimating the degree of ionization in optically thick
absorption systems.

     Our simulations predict a uniform layer 
with $\N{H^0} = 2.4 \sci{20} \cm 2$ and
$n_H = 0.1 \; \rm cm^{-3}$ will be $96\%$ neutral.  
The results should be similar for
a more realistic multicomponent system, since we expect the majority
of the 23 components in the $z$=2.46 system to be shielded 
from ionizing radiation by gas
both above and below the location of the 
cloud in the layer, thereby maintaining
neutrality (Figure 14b).  Therefore, we measure the metal abundances of
our damped \Lya system by assuming the
abundance of element X to equal $\N{X^+} / \N{H^o}$, where $X^+$ is the first
ionization state of element X.  Note, however, that 
this result is very dependent
on the value of $n_H$.  For instance, 
we find a neutral fraction of $\approx 37\%$ 
for $n_H = 10^{-3} \; \rm cm^{-3}$ and $\N{H^0} = 2.4 \sci{20} \cm2$.
Of course, physically this would be imply a very large system
with dimensions exceeding 100 kpc.

\subsubsection{CLOUDY simulations}

     We have performed CLOUDY photoionization simulations with the aim of
using the relative abundances of different 
ionization states of the same element to
estimate the degree of ionization of our system.  The 
analysis parallels the
treatment presented by Lu et al. (1995).  For the input radiation 
we used an
attenuated power law ionizing spectrum 
computed by Madau (1992).  The spectrum was
normalized to a mean intensity of $J_\nu = .195 \sci{-21} \>
\rm ergs \> s^{-1} \cm2 \> Hz^{-1} \> sr^{-1}$ at 1 Rydberg calculated at an
average redshift of $z$=2.46.  
The calculations assume a plane-parallel geometry
with only one face illuminated and with incident 
radiation perpendicular to the
surface, in contrast to the more realistic assumptions used to
calculate H$^o$/H in the previous
section.  The metallicity is fixed at [Z/H] = $-0.5$, 
the hydrogen volume density is
varied, and the program terminates at a column density  $\log [\N{H^o}] = 20.
38$.  Thus the neutral hydrogen column density is 
fixed at the observed value $\log [\N{H^o}] = 20.38$
while the total hydrogen column density is allowed to vary.  Although this
simulation is overly simplified 
(e.g. single side illumination, normal incidence, no
dust depletion model), it does provide a good estimate 
of the degree of ionization.
Figure~\ref{I-Comp} presents the results of the simulations.  
Note that no conclusions
concerning the degree of ionization can be drawn by comparing the relative
abundances of the low-ions because their column 
densities track $\N{H^0}$ which is held constant.

        Given the CLOUDY results, we can use the observed
 Si$^+$ to Si$^{3+}$ and Al$^+$ to Al$^{++}$ ratios to  determine
the ionization level of the gas.  Since the AlII profile is heavily
saturated, we take
$N(v) = 12.00$ per pixel corresponding to $I_a(v)/I_i(v) = 0.05$
in the optical depth equation (Equation 1).  Given the degree of saturation,
this
value yields a {\em very conservative} lower limit for [Al$^+$/Al$^{++}$]
$\equiv \log[\N{Al^+}] - \log[\N{Al^{++}}]$.  Figure~\ref{Rtio}
is a plot of (a) [Si$^+$/Si$^{3+}$] and (b) [Al$^+$/Al$^{++}$]
for 5 pixel bins over the entire low-ion region.  The large
dots represent the average value over the velocity regions defined
by the vertical dashed lines and the borders of the plot.
Table~\ref{rtio_tab} gives the
values of [Si$^+$/Si$^{3+}$] and [Al$^+$/Al$^{++}$]
as well as the corresponding H$^0$/H$^+$ ratio derived from the CLOUDY
results for the velocity regions.

        With the exception of the region $-40$ \kms $< v < $ 40 \kms,
we find [Al$^+$/Al$^{++}$] $>$ 0.4 dex.  This yields a neutral
hydrogen fraction H$^0$/H $>$ 0.5. The 
velocity region $v$ = [-100, 40] {\kms} corresponds
to the strongest feature in all of the low-ion profiles and we expect
the abundance of Al II to greatly exceed the value assumed above.
The [Si$^+$/Si$^{3+}$]
results are more difficult to interpret because the Si IV line-profile
does not closely trace the low-ion or Al III profiles.  According to
the SiII to SiIV ratio, the most highly ionized region 
is $-218$ \kms $ < v < - 140$ \kms
which is in clear contradiction with the [Al$^+$/Al$^{++}$] results.
We believe that a significant portion of the SiIV absorption is
due to gas which is physically separate from the low-ion and Al$^{++}$ gas.
As such, we consider the [Si$^+$/Si$^{3+}$] values to be lower limits.
To summarize, our results suggest
the hydrogen fraction of the gas
H$^0$/H $>$ 0.5 is conservative lower limit. Therefore, it is highly likely
that the gas in the $z$ = 2.46 damped system is mainly neutral.

It is also likely that the neutrality of the gas is unaffected by the
presence of {\em internal} sources of ionizing radiation such as O B stars.
We checked this possibility by performing CLOUDY calculations with
input from a black body with T = 4$\times$10$^{4}$ K. The resulting
Al$^{+}$/Al$^{++}$ and Si$^{+}$/Si$^{++}$ ratios are indistinguishable
from the ratios predicted by the external radiation field considered
above. The results are similar because the sharp drop
in ionizing flux at the Lyman limit in the attenuated external spectrum
(\cite{mad92}) mimics the exponential fall off of a black body. As a result
the metal-line ratios indicate H is mainly neutral in the $z$ = 2.46
absorber, for plausible sources of external and internal radiation.

\section{RESULTS}

     This section presents the 
results from our abundance and kinematic analyses
of the damped \Lya system at $z$=2.462.  We discuss the evidence for dust
in this system and remark on the nearly constant abundances relative
to Zn for three of the velocity features in our system.  
Finally, we compare and
contrast the kinematics of the damped \Lya system with several other systems,
including another published HIRES damped \Lya system (\cite{wol94}).

\subsection{Abundances of the $z$=2.462 System}

     Table~\ref{abnd} lists the column density 
$\log[\N{X}]$, and the logarithmic
abundance of element X relative to hydrogen normalized to solar abundances,
[X/H] $\equiv \log[\N{X}/\N{H}] - \log[\N{X}/\N{H}]_\odot$, for the
damped \Lya system at $z$=2.462.  The abundance is derived assuming
$\log [\N{H}] = 20.38 \pm 0.045$ (10$\%$ error)
and standard solar
abundances (\cite{and89}).  Our results indicate a relatively
metal-rich system.  If the three Zn$^+$ features (discussed in $\S 5.1$)
comprise 50$\%$ of the total
Zn$^+$ abundance (an analysis of these components in the other low-ions
suggests $\approx 45\%$), we find [Zn/H] $= -0.262$. 
This is the most metal-rich
of any damped \Lya system for which accurate abundances have been determined
at redshift $z \geq 2.0$.   
Furthermore, it has a higher metallicity (assuming Zn predicts metallicity;
see below) than all but
one of the systems measured in the most extensive survey of metallicity in
damped \Lya systems carried out so far (\cite{ptt94}).

          Figure~\ref{RelAbd} and Table~\ref{depl} present the abundances
of the low-ions relative to Zn (assuming cosmic abundances) 
in the three velocity features where we could reliably
measure the Zn abundance.  The error bars are not entirely 
accurate (they are derived with the apparent optical depth method), 
but do serve as valuable guides.
The overall variation of the abundances relative to Zn is very small for
the 3 features and all of the elements are depleted with respect to Zn.  
The values of [X/Zn], however, do show a slight increase 
from the first to the third feature.
We believe the most likely explanation lies in our measurements of the Zn
abundance, particularly since features 1 and 2 are more 
likely to be blended with
absorption from Cr II 2062 and because the trend is 
relatively systematic.  Even
given these minor differences, the relative abundances between Fe,
Ni, Si, Cr and Zn are essentially constant over 
the 3 features (corresponding to nearly 200 \kms in velocity space).  
Therefore, we see little evidence for
gas-phase abundance variations throughout our system.  
This observation indicates
damped \Lya systems are chemically well-mixed which 
further suggests they are detected
at ages large compared to their internal dynamical time scales 
(i.e. the rotation period).

\subsection{Dust Depletion}

        In this section we analyze
variations of the gas-phase element abundances
in damped {\lya} systems with (a) velocity and (b) condensation
temperature. Such variations have been observed in the ISM of the Galaxy
and have been used to infer properties of the dust responsible for
element depletion. We wish to see whether similar effects are
present at high redshifts.

\subsubsection{Abundance Variations}

The absence of variations in  [X/Zn]
with respect to velocity
in the $z$ = 2.462 system is contrary to the presence
of such variations in the ISM.  Spitzer
\& Fitzpatrick (1993) and Spitzer \& Fitzpatrick (1995) 
recently used GHRS spectra to
infer these variations for the sightlines to
the Galactic stars HD 93521 and HD 149881. These
sightlines are relevant because the measured $N$(H$^{0}$) are similar to the
value inferred for the damped system. Moreover,  the depletion levels
are similar, {\em provided one interprets negative 
values of {\rm [X/Zn]} in damped \lya systems to result
from grain depletion.}
 From Figure~\ref{RelAbd} and Table~\ref{depl} we infer
an upper limit of $\approx$ 0.3 dex for the variation
of [Si/Zn], [Fe/Zn], [Ni/Zn], or [Cr/Zn] across the 3 velocity features
in the $z$ = 2.46 damped system, while the
variation for [Fe/Zn] in HD 93521
is $\approx$ 0.9 dex and for [Cr/Zn] in HD 149881 is 0.6 
dex (\cite{spz93,spz95}).
Fitzpatrick and Spitzer attribute the changes in [X/Zn] to
the increase in dust destruction for clouds with increasing random
velocity with respect to galactic rotation speed.

Variations in [X/Zn] could also be due to density variations along
the line of sight.
Studies of the ISM (\cite{jen87}) have shown correlations
between the average Hydrogen volume density $n_H$ and the degree of depletion
$D$ relative to cosmic abundances for a variety of elements,
including Si, Cr, and
Fe. Jenkins fitted the ISM data with the following depletion curve:

\begin{equation}
D = d_0 + m \, \left [ \, \log n_H + 0.5 \, \right ]
\end{equation}

\noindent where $D$ is the logarithmic depletion relative to cosmic
abundances (i.e., [X/H]),
$m$ is the slope, $d_0$ is the value of 
$D$ at log $n_H = -0.5$ and $n_H$ is the
Hydrogen volume density averaged 
over the line of sight to a given star.  It is
straightforward to show that the difference of two depletion
values corresponding
to two measurements ($\Delta D \equiv D_2 - D_1$) can be related to
the ratio of the $n_H$ values of each measurement:

\begin{equation}
{(n_H)_2 \over (n_H)_1} = 10^{\big | {\Delta D \over m} \big |} \quad .
\end{equation}

        As stated above
the variation of Si, Cr, and Fe
relative to Zn  over the
3 features is only $\approx$ 0.3 dex, 
where the largest variation is between the
first and third features (the reddest and bluest).
As suggested in $\S 7.1$ this variation is almost certainly systematic,
most likely a result of blending with Cr II 2062.  Therefore, the value
can be considered an \underline{upper limit}
to the actual abundance variation over
these 3 components.  Table~\ref{nH} lists
the depletion variations, the ISM slopes cited in Jenkins (1987),
the corresponding predicted $n_H$ variation, 
and 1$\sigma$ errors for Fe, Si, and
Cr.  We note the Cr measurements place the tightest constraints on the
predicted variation of $n_H$. 
Of course, the errors are rather large, but we can
confidently predict that the variations of $n_H$ are 
no larger than $\approx 1$
dex
provided this technique is applicable.

What suppresses variations of [X/Zn] in the damped system?
If dust is present, then the sightline through the damped system must
encounter an ISM in which $n_{H}$ varies by less than a factor of 10
along the line of sight. Moreover, the efficiency of grain destruction
by shocks must be less efficient than in the ISM. Because the frequency
of supernovae explosions account for the inhomogeneous nature of the
ISM (\cite{mck77}) and  because supernovae are the main
contributor of shocks in the ISM, then a lower frequency of supernova
explosions could account for the lack of variations in the damped system.
However, in the ISM a sightline of a few kpc typically
encounters 4 orders of magnitude variation in $n_{H}$ (\cite{kul87}).
A drastic reduction
in the rate of supernova explosions would be required to reduce this
to one order  of magnitude. But models for the evolution of disk galaxies
suggest that supernova input in the past was stronger not weaker
than in the present (\cite{bru92}). Other, nucleosynthetic considerations
suggest that  supernovae rates in the past were as least as
frequent as they are at present (\cite{tru95}).
Therefore, while we cannot rule out this explanation
altogether, it seems implausible.

Therefore, the null detection  of variations in [X/Zn] requires
a weaker coupling of the depletion rate to the gas density, and
a lower grain destruction rate by interstellar shocks than in
the ISM. As a result, 
dust in the damped system is either absent or
has significantly different properties from dust in the ISM.

\subsubsection{Condensation Temperature}

     Figure~\ref{Tcond} plots the gas phase abundances versus 
condensation temperature, $T_C$, for the observed
low-ions from the damped \Lya system (solid dots) 
and from the line-of-sight toward the Galactic star $\zeta$ Oph
(\cite{crd94}; open squares\footnote{The error bars for
the ISM data are on the order of the size of the squares or smaller.}).
The idea is to
determine whether the anti-correlation between [X/H] and $T_C$ observed in
the ISM (\cite{jen87}) is also present in high-$z$ damped 
\Lya systems.  Because
$T_C$ is the temperature at which half the gas phase atoms in a stellar
atmosphere condense to solid form, the ISM anti-correlation is taken
as evidence for dust formation in stellar atmospheres (\cite{fie74}).
Comparison between the ISM and damped diagrams shows some similarities.
Specifically, the relative rankings of [Cr/H], [Si/H], [Fe/H], and
[Ni/H] are about the same.  However, the absolute values are significantly
higher in the damped system. In addition [Zn/H] in the ISM greatly
exceeds all the other [X/H], but in the damped system [Zn/H] does
not greatly exceed [Cr/H], [Fe/H], and [Ni/H], and may in fact
be comparable to [Si/H]. The differences between the ISM and damped
[X/H] may indicate the presence of gas in which the metallicity
(inferred from
undepleted [Zn/H]) is $> -0.5$ and has a lower
dust-to-gas ratio to account for the smaller difference between
[Zn/H] and [X/H] (cf. \cite{ptt94}).   

	On the other hand, Lu et al.\ (1995, 1996a, 1996b) interpret
these patterns in terms of nucleosynthetic yields from type II
supernovae.  They point out that the damped \lya abundance patterns
of N/O, Si/Fe, Cr/Fe, and Mn/Fe are consistent with the abundance
patterns observed for population II halo stars which have
been primarily enriched by type II supernovae (\cite{whe89}).
They also stress that Mn/Fe for damped \lya systems and halo stars
disagrees with Mn/Fe inferred for the ISM, which is influenced
by grain depletion.  The only difficulties with the Lu et al.\
hypothesis stem from Zn/Fe.  The quantity [Zn/Fe] $>$ 0 in damped
\lya systems, while [Zn/Fe] $\approx 0$ for stars with
$-3 <$ [Fe/H] $< 0$ (\cite{sne91}).  

Theorists working on chemical
evolution have long been puzzled as to why [Zn/Fe] is independent
of [Fe/H].  The closed box model by Malaney and Chaboyer (1996)
predicts [Zn/Fe] to increase with [Fe/H] owing to the 
metallicity-dependent yield computed for type II supernovae explosions.
These authors suggest that the constant [Zn/Fe] could be peculiar
to the chemical evolution of the Galaxy, due to non-LTE effects in 
stellar atmospheres, or errors in the calculated yields.   In 
addition, while the Malaney-Chaboyer model predicts [Zn/Fe] $< 0$,
Hoffman et al.\ (1996) suggest neutrino driven winds in type II
supernovae might result in [Zn/Fe] $> 0$.  This effect may help explain
the overabundance of Zn relative to Fe in damped \lya systems.
Therefore, while some
questions remain to be addressed, nucleosynthetic yields and grain
depletion are equally plausible explanations for the abundance
patterns observed in damped \lya systems.

	These competing explanations lead to at least two different
ways for interpreting
the metallicity of damped \lya systems.  First, if the Lu et al.\
interpretation is correct and Zn is somehow overproduced
with respect to Fe, than the metallicity of damped \lya systems
is significantly lower ($\approx 0.7$ dex) than estimated from
the Zn abundances.  In terms of the system analyzed in this paper,
the metallicity would be $-0.83$ dex which would still be the
highest metallicity observed in a damped \lya system with $z > 2.0$.
On the other hand, if we accept depletion, the metallicity of these
systems is at least as high as the Zn abundance, and 
the Malaney-Chaboyer results indicate it could 
be even higher (possibly requiring an even
greater level of depletion).

\subsection{Kinematics of the Damped \Lya System ($z$=2.462)}

     In this subsection we discuss the kinematics of 
the damped \Lya system at
$z$=2.462. We stress again the observed differences between the high 
and low-ion
profiles and intercompare the velocity profiles of the $z$=2.462 system and
several other damped systems.

\subsubsection{Comparison of the Low and High Ion Profiles}

     Although all of the transitions from the $z = 2.462$ system
span approximately the same
velocity interval ($\approx$ 200 \kms), there are marked
differences between the low and high-ion line profiles.  Similar
to the damped \Lya system toward PHL 957 discussed by 
Wolfe et al. (1994), the
velocity profiles of all of the low-ions are visibly asymmetric
with the component at the red edge being the strongest of the profile.
In this case, the strongest component is at the red edge, where 
in PHL957 the
strongest component is at the blue edge.
Figure~\ref{Per-fig} and Table~\ref{per} give the column density
by percent of total in 5 binned intervals in
velocity space corresponding to visible features in the line profiles.
This serves as quantitative evidence of the asymmetry in the low-ions
and highlights the inherent differences between the low and high-ion
profiles.  The overall profile of the high-ions has nearly 
the opposite shape of the
low-ion profiles with the strongest feature on the blue edge.  In
addition, the profile of the high-ions is smoother with less 
structure than the low ions.

     We contend these differences, particularly those evident in
the Si II and Si IV profiles, indicate the absorption associated with the
high-ions does not originate in the same region as that
associated with the low-ions.  The most compelling evidence for
this hypothesis lies in the extreme difficulty encountered in obtaining an
accurate fit to the Si IV profile by altering the column densities
of the calculated Si II components while holding the redshift and
$b$ values constant.  Both the smoothness and the overall shape of
the Si IV profile require significantly different $b$ values and
argue against the inclusion of so many thin, resolved components.

\subsubsection{The $z$=2.457 subsystem}

     The neighboring system at $z$=2.457 was unidentified
in the lower resolution data.  It has similar line profile structure as the
$z$=2.462 system with a large velocity interval, smoother high-ion profiles
and low-ion profiles with more velocity structure.  It does not, 
however, exhibit
the edge leading asymmetry apparent in the $z$=2.462 system.  In fact, the
low-ions appear relatively symmetric about $v = 70$ \kms.  We contend these
differences in the kinematic characteristics result from the fact that the
$z$=2.457 system is not a damped \Lya system but most likely an ionized 
Lyman limit
system in which the gas kinematics are not determined by rotation.

\subsubsection{Comparison with PHL 957}

     Contrasting the damped \Lya metal transitions with those
toward PHL 957 at $z$ = 2.309, we note several important
differences.  First, the velocity interval of the Q0201+365 
damped system is
significantly larger (200 \kms vs. 50 \kms).  Secondly, the low-ions of
the Q0201+365 system exhibit more velocity structure, an effect 
which is not
due to differences in resolution.   Finally, the asymmetric
shape is not as prevalent in our damped system possibly because 
of its larger
velocity interval.  That is, the gas is more evenly distributed 
in Q0201+365
than in the PHL957 absorber.  In terms of the thick rotating 
disk model, these
differences may be explained by differences in 
the inclination angle
of the line-of-sight with respect to the disk.

\subsubsection{Comparison with System at $z$ = 2.325}

      A comparison of the damped \Lya kinematics with the
absorption system at $z$ = 2.325 toward Q0201+365 further stresses the
damped \Lya characteristics. The $z$ = 2.325 system (a possible Lyman limit
system) also spans $\approx$ 200 \kms in velocity space, but does not possess
the same edge-leading asymmetric profile observed in the damped
systems.  Furthermore, the high-ion profiles at $z$=2.325 trace
the low-ion profiles more closely than in the damped \Lya
systems. For instance, Al III was successfully tied to the low-ion 
VPFIT solution.
These characteristics are all consistent with an explanation depicting the 
$z$=2.325
system as a multi-cloud system in which the velocities are random 
rather than systematic
as in the case of rotation.  Moreover the gas is more evenly
distributed.  In short, the $z=2.325$ system more 
closely resembles the $z$=2.457 system
which is also a likely Lyman limit system.  
Perhaps we are observing kinematic
characteristics which differentiate damped \Lya from Lyman limit systems.  
However, the statistics of small numbers makes this observation a 
speculation rather than a conclusion.

\section{CONCLUSIONS}

     This paper presented HIRES spectra obtained with the Keck 
10m telescope of 
absorbing gas toward toward Q0201+365.  We identified over $80\%$ of the 
absorption features
and have analyzed several of the more interesting metal-line systems.
We have focused on the damped \Lya system at $z$=2.462 as part of an ongoing
program to investigate the chemical content and kinematics of damped systems
within the redshift interval $z \approx 2 - 4$.
We summarize our results as follows.

(1)  Based on the analysis of ionization simulations, we predict
the damped \Lya system to be significantly neutral.  Although it is
possible the system is partially ionized, our analysis predicts the metals
are all essentially in the singly ionized state, and that
the total hydrogen column density is well within a factor of two of
the adopted value from the $\N{H^0}$ measurement. 

(2) A hidden component analysis of the Ni II 1741, 1751 transitions did not
reveal any significant hidden saturated components.  We expect this
to hold true for the other low-ion transitions.  

(3)  With the VPFIT least squares line profile fitting
package we have measured ionic column densities for the damped \Lya
system at $z$=2.462 (as well as several other systems). We performed similar
measurements with the apparent optical depth method and found the two results
to be in agreement, further eliminating  the possibility of hidden
line saturation.

(4)  We measured the following abundances of Si, Fe, Cr, and Ni for the damped
\Lya system: [Si/H] = $-0.376 \pm 0.052$, [Fe/H] = $-0.830 \pm 0.051$,
[Cr/H] = $-0.902 \pm 0.064$, and [Ni/H] = $-1.002 \pm 0.054$. 
We placed limiting values on the abundances of the s-process elements
Pb (3$\sigma$ detection) and Ge (upper limit), [Pb/H] = $2.233 \pm 0.121$ 
and [Ge/H] $<$ 0.664,  and a lower limit value on the abundance of 
Zn, [Zn/H] $> -0.562$.  Based on the VPFIT solution
of the low-ions, we expect the metallicity is [Z/H] $\approx -0.262$.  
This damped \Lya system has the highest metallicity measured to 
date at $z \geq 2.0$.

(5)  Comparing individual features of the damped \Lya system, we find the
relative abundance between Si, Fe, Cr, Ni and Zn remains nearly constant
throughout our system (Figure~\ref{depl}).  This suggests a well mixed system
with an age large compared to the internal dynamical time scale at the epoch
of detection.

(6)   We have used the relatively minor variations 
observed in the Si, Cr, and
Fe abundances relative to Zn to place limits on the expected variation in the
Hydrogen volume density throughout the damped \Lya system, having assumed the
presence of dust grains and the Jenkins relation (\cite{jen87}).  Our
measurements of [Cr/Zn] place a maximum variation of $n_H$ at $\approx 1$ dex.
The lack of $n_H$ (and [X/Zn]) variations could be evidence of weaker supernova
input in the past, but we believe they are more likely due to the absence
of grains with the properties of dust found in the ISM.

(7)  Plotting the measured abundances versus condensation temperature 
(Figure~\ref{Tcond}),
we do find evidence for a depletion pattern, but the overall depletion level
of Si, Fe, Cr and Ni with respect to Zn is indicative of 
a relatively dust free ISM cloud.  Although gas
with a lower dust-to-gas ratio than evident in the ISM can account
for the pattern, one can also explain
the pattern in terms of nucleosynthetic yields from type II supernovae
(\cite{lu95b}).  Both explanations are problematic and are under
continued debate.  Determining the proper explanation is particularly
important as they predict different metallicities
for damped \lya systems, which will significantly affect the investigation
of galactic chemical evolution.

(8) The low-ion profiles of the damped system exhibit an edge leading 
asymmetry as predicted by a simple model of rotation.  The shape is similar
to the other damped system observed with HIRES (PHL 957; \cite{wol94}),
though the velocity interval is significantly greater.

\acknowledgments

The authors would like to thank Bob Carswell for providing the line-profile
fitting package VPFIT as well as Tom Barlow for his excellent HIRES data 
reduction software.  
We would also like to thank Vincent Virgilio for his help
in developing the neutral hydrogen model and Piero Madau for
providing the ionizing spectrum.  Finally, we would like to thank
Ed Jenkins and Edward Fitzpatrick for helpful discussions.
AMW and JXP were partially supported by 
NASA grant NAGW-2119 and NSF grant AST 86-9420443.  

\clearpage

\begin{table*}
\begin{center}
\begin{tabular}{lccc}
UT Date
& Exposure
& Wavelength
& Resolution\\ 
& Time (s) & Coverage ($\rm \rAA$) & (\kms) \\
\tableline
1994 Sep 15 & 9600 & 4720 - 7130 & 7.2 - 8.0 \cr
1994 Sep 30 & 11950 & 4790 - 7180 & 7.0 - 8.0 \cr
1994 Oct 1 & 13030 & 4790 - 7180 & 7.2 - 8.1 \cr
\end{tabular}
\end{center}

\caption{JOURNAL OF OBSERVATIONS} \label{obs}

\end{table*}

\clearpage

\begin{deluxetable}{lcccll}
\tablewidth{0pc}
\tablenum{2}
\tablecaption{ABSORPTION LINE LIST} 
\tablehead{
\colhead{Order} & \colhead{$\lambda$} &
\colhead{W} & \colhead{$\sigma_W$} &
\colhead{ID} & \colhead{$z_{abs}$} \nl
& \colhead{$(\rAA)$} & 
\colhead{$(\rAA)$}} 

\tiny
\startdata
75  \nl
& 4724.79 & 0.1891 &  0.0101 &           &       \nl
& 4727.73 & 0.1681 &  0.0089 &           &       \nl
& 4730.96 & 0.0456 &  0.0049 &           &       \nl
& 4737.23 & 0.2245 &  0.0081 &           &       \nl
& 4739.54 & 0.3687 &  0.0076 &           &       \nl
& 4742.94 & 0.3541 &  0.0132 & NiII(1370)  & 2.462  \nl
& 4750.76 & 0.1077 &  0.0062 &           &       \nl
& 4753.96 & 1.8529 &  0.0125 & FeII(1608)  & 1.956 \nl
& 4758.58 & 1.6854 &  0.0073 &              &       \nl
& 4761.95 & 0.0246 &  0.0038 & FeII(1611)  & 1.956  \nl
& 4763.30 & 0.0170 &  0.0031 &           &       \nl
& 4768.05 & 0.0344 &  0.0040 & SiIV(1393)  & 2.421 \nl
& 4770.63 & 0.2802 &  0.0067 & SiIV(1393)  & 2.423 \nl
& 4772.64  & B\tablenotemark{a} & & SiIV(1393)  & 2.424 \nl
& 4773.34 & B\tablenotemark{a} &            &       \nl
& 4783.89 & 0.9927 &  0.0099 &              &       \nl
74  \nl
& 4784.71 & 0.2125 &  0.0053 &           &       \nl
& 4798.95 & 0.0257 &  0.0029 & SiIV(1402)  & 2.421 \nl
& 4802.23 & 0.1792 &  0.0082 & SiIV(1402)  & 2.423 \nl
& 4808.08 & 0.0957 &  0.0046 &           &       \nl
& 4809.61 & 0.0474 &  0.0029 &       &       \nl
& 4817.94 & 1.5530 &  0.0104 & SiIV(1393)  & 2.457 \nl
& 4824.06 & 1.9596 &  0.0105 & SiIV(1393)  & 2.462 \nl
& 4842.24 & 1.6780 &  0.0067 & MgII(2796)  & 0.732 \nl
& 4847.31 & B\tablenotemark{a} & & MgII(2796)  & 0.733 \nl
& 4849.94 & B\tablenotemark{a} & & SiIV(1402)  & 2.457 \nl
& 4852.51 & 0.0287 &  0.0052 &           &       \nl
73  \nl
& 4850.48 & 0.3522 &  0.0067 & SiIV(1402)  & 2.457 \nl
& 4855    & B\tablenotemark{a} & & MgII(2803)  & 0.731 \nl
& 4855    & B\tablenotemark{a} & & SiIV(1402)  & 2.462 \nl
& 4859.84 & 0.2065 &  0.0061 & MgII(2803)  & 0.733 \nl
& 4863.60 & 0.0227 &  0.0026 &           &       \nl
& 4876.08 & 0.0190 &  0.0028 &           &       \nl
& 4895.98 & 0.0982 &  0.0087 & CI(1656)    & 1.955 \nl
72  \nl
& 4924.10 & 0.1703 &  0.0073 & CII(1334)   & 2.690  \nl
& 4938     & B\tablenotemark{a} & & AlII(1671)  & 1.955 \nl
& 4940     & B\tablenotemark{a} & & MgI(2852)   & 0.732 \nl
& 4951.84 & 0.2428 &  0.0112 & SiIV(1393)  & 2.553  \nl
& 4984.16 & 0.2462 &  0.0152 & SiIV(1402)  & 2.553  \nl
71  \nl
& 5005.84 & 1.3215 &  0.0112 & CIV(1548)   & 2.233 \nl
& 5014.00 & 0.8942 &  0.0101 & CIV(1550)   & 2.233 \nl
70  \nl
& 5076.49 & 1.3223 &  0.0153 & SiII(1526)  & 2.325 \nl
69  \nl
& 5139.61 & 0.3223 &  0.0059 & CIV(1548)   & 2.320 \nl
& 5142.70 & 0.7369 &  0.0094 & SiIV(1393)  & 2.690  \nl
& 5148     & B\tablenotemark{a} & & CIV(1548)   & 2.325 \nl
& 5148.15    & B\tablenotemark{a} & & CIV(1550)   & 2.319 \nl
& 5156.16 & 2.0353 &  0.0133 & CIV(1550)   & 2.325 \nl
& 5172.86    & S\tablenotemark{b} & & Sky Emiss   &       \nl
& 5175.62 & 0.8012 &  0.0136 & SiIV(1402)  & 2.690  \nl
& 5177.77 & 0.0241 &  0.0035 &           &       \nl
& 5189.73 & 0.0407 &  0.0054 & CI(1560)    & 2.325 \nl
& 5200.18 & 0.0301 &  0.0054 & Sky??       &       \nl
& 5201.43 & 0.0609 &  0.0057 & CIV(1548)   & 2.360 \nl
\tablebreak
68  \nl
& 5210.07 & 0.0330 &  0.0047 & CIV(1550)   & 2.360 \nl
& 5234.86 & 0.0337 &  0.0049 & Sky Abs     &       \nl
& 5270.54 & 0.1121 &  0.0076 &        &       \nl
& 5272.51 & 0.2489 &  0.0128 &        &       \nl
& 5275.32 & 0.1389 &  0.0077 &        &       \nl
& 5278.45 & 1.5553 &  0.0177 & SiII(1526)  & 2.457 \nl
67  \nl
& 5285.95 & 3.3202 &  0.0121 & SiII(1526)  & 2.462 \nl
& 5298.86 & 2.4749 &  0.0171 & CIV(1548)   & 2.423 \nl
& 5307.77 & 1.7411 &  0.0164 & CIV(1550)   & 2.423 \nl
& 5326.05 & 0.3089 &  0.0069 & SiIV(1393)  & 2.821 \nl
& 5342.40 & 0.2902 &  0.0131 & SiII(1808)  & 1.955 \nl
& 5347.39 & 0.2201 &  0.0148 & FeII(1608)  & 2.325 \nl
& 5352.75 & 3.2158 &  0.0166 & CIV(1548)   & 2.457 \nl
& 5357.45 & 2.3480 &  0.0169 & CIV(1548)   & 2.462 \nl
66  \nl
& 5367.53 & 2.1787 &  0.0194 & CIV(1550)   & 2.462 \nl
& 5373.05 & 0.2364 &  0.0091 & CIV(1550)?  & 2.465  \nl
& 5383.04 & 0.3361 &  0.0130 & CIV(1548)   & 2.477 \nl
& 5391.61 & 0.1891 &  0.0124 & CIV(1550)   & 2.477 \nl
& 5419.45 & 0.4062 &  0.0097 & CIV(1548)   & 2.500 \nl
& 5425.90    & S\tablenotemark{b} & & Sky Abs     &       \nl
& 5428.50 & 0.2452 &  0.0112 & CIV(1550)   & 2.500 \nl
65  \nl
& 5458.78 & 0.0286 &  0.0039 &           &       \nl
& 5481.60 & 1.3494 &  0.0128 & AlIII(1854) & 1.955 \nl
& 5500.98 & 1.5163 &  0.0100 & CIV(1548)   & 2.553  \nl
& 5505.02 & 0.7522 &  0.0135 & AlIII(1862) & 1.955 \nl
& 5510.20 & 0.9414 &  0.0119 & CIV(1550)   & 2.553  \nl
& 5512.77 & 0.0255 &  0.0042 &           &       \nl
64  \nl
& 5555.32 & 1.3290 &  0.0144 & AlII(1671)  & 2.325 \nl
& 5560.52 & 0.2871 &  0.0130 & FeII(1608)  & 2.457 \nl
& 5568.31 & 2.6714 &  0.0117 & FeII(1608)  & 2.462 \nl
& 5579.22    & B\tablenotemark{a} & & FeII(1611)  &       2.462 \nl
& 5607.50 & 0.0793 &  0.0079 &           &       \nl
63  \nl
& 5622.07 & 0.0397 &  0.0062 &           &       \nl
62  \nl
& 5713.13 & 1.0943 &  0.0136 & CIV(1548)   & 2.690 \nl
& 5722.31 & 0.6010 &  0.0125 & CIV(1550)   & 2.690 \nl
& 5776.46 & 2.1480 &  0.0137 & AlII(1671)  & 2.457 \nl
& 5783.40 & 5.0321 &  0.0164 & AlII(1671)  & 2.462 \nl
61  \nl
& 5804.51 & 0.5622 &  0.0091 & FeII(2344)  & 1.476 \nl
& 5845.15 & 0.0226 &  0.0044 &           &       \nl
& 5872.81 & 0.0190 &  0.0037 & FeII(2374)  & 1.473 \nl
& 5874.03 & 0.1010 &  0.0072 & FeII(2374)  & 1.474 \nl
& 5879.60 & 0.2519 &  0.0072 & FeII(2374)  & 1.476 \nl
60  \nl
& 5901.21 & 0.4404 &  0.0099 & FeII(2382)  & 1.476 \nl
& 5916.32    & B\tablenotemark{a} & & CIV(1548)?  & 2.821 \nl
& 5919     & B\tablenotemark{a} & & NiII(1709)  & 2.462 \nl
& 6076.31 & 0.0191 &  0.0033 & CrII(2056)  & 1.955 \nl
59  \nl
& 6010.37 & 0.0947 &  0.0127 & SiII(1808)  & 2.325 \nl
& 6028.97 & 0.3928 &  0.0113 & NiII(1741)  & 2.462 \nl
& 6043.72    & S\tablenotemark{b} & & Ink Blot    &       \nl
& 6064.91 & 0.2400 &  0.0110 & NiII(1751)  & 2.462 \nl
\tablebreak
58  \nl
& 6166.72 & 0.6211 &  0.0140 & AlIII(1854) & 2.325 \nl
57  \nl
& 6259.01 & 0.6740 &  0.0140 & SiII(1808)  & 2.462 \nl
& 6280 -   & S\tablenotemark{b} & &  Sky Abs    &       \nl
56  \nl
& 6327.85 & 0.3591 &  0.0090 & FeII(2344)  & 1.699 \nl
& 6331.86    & S\tablenotemark{b} & & Sky Abs    &       \nl
& 6366     & S\tablenotemark{b} & & Sky Abs    &       \nl
& 6380.61    & S\tablenotemark{b} & & Sky Abs   &       \nl
& 6398.86 & 0.2889 &  0.0095 & FeII(2586)  & 1.473 \nl
& 6400.33 & 0.0246 &  0.0040 & FeII(2586)  & 1.474 \nl
& 6403.12 & 0.0902 &  0.0061 & FeII(2586)  & 1.475 \nl
& 6404.76 & 0.5519 &  0.0089 & FeII(2586)  & 1.476 \nl
& 6411.14 & 0.2570 &  0.0113 & FeII(2374)  & 1.699 \nl
55  \nl
& 6438.10 & 0.9346 &  0.0102 & FeII(2600)  & 1.476 \nl
& 6440.63 & 0.2097 &  0.0109 & AlIII(1863) & 2.457 \nl
& 6448.40 & 0.9757 &  0.0148 & AlIII(1863) & 2.462 \nl
& 6470 -   & S\tablenotemark{b} & & Sky Abs   &       \nl
54  \nl
53  \nl
52  \nl
& 6813.84 & 0.0381 &  0.0053 &          &       \nl
& 6869 -   & S\tablenotemark{b} & & Band Abs    &       \nl
51  \nl
& 6942.03 & 0.9711 &  0.0084 & MgII(2803)  & 1.476 \nl
& 6944.50    & S\tablenotemark{b} & & Sky Abs     &       \nl
& 6946.10    & S\tablenotemark{b} & & Sky Abs     &       \nl
& 6958.54    & S\tablenotemark{b} & & Sky Abs     &       \nl
& 6963.62    & S\tablenotemark{b} & & Sky Abs     &       \nl
& 6982.78 & 0.2482 &  0.0105 & FeII(2586)  & 1.699 \nl
& 6987.39    & S\tablenotemark{b} & & Sky Abs     &       \nl
& 6988.89    & S\tablenotemark{b} & & Sky Abs     &       \nl
& 6991.29    & S\tablenotemark{b} & & Sky Abs     &       \nl
& 6995.81    & S\tablenotemark{b} & & Sky Abs     &       \nl
& 7001.26    & S\tablenotemark{b} & & Sky Abs     &       \nl
& 7007.03    & S\tablenotemark{b} & & Sky Abs     &       \nl
& 7016     & B\tablenotemark{a} & & FeII(2374)  & 1.955 \nl
& 7018.78  & B\tablenotemark{a} & & FeII(2600)  & 1.699 \nl
& 7039.67 & 1.6433 &  0.0080 & FeII(2382)  & 1.955 \nl
50  \nl
& 7118.18 & 0.2471 &  0.0159 & CrII(2056)  & 2.462 \nl
& 7140.5   & B\tablenotemark{a} & & CrII(2062)  & 2.462 \nl
& 7140.5   & B\tablenotemark{a} & & ZnII(2062)  & 2.462 \nl
& 7144 -   & S\tablenotemark{b} & & H$_2$O Abs  &       \nl
\tablenotetext{a}{ Line blending}
\tablenotetext{b}{ Night-sky absorption or emission features}
\enddata
\normalsize
\end{deluxetable}

\begin{deluxetable}{lcc}
\tablewidth{0pc}
\tablenum{3} 
\tablecaption{METAL TRANSITIONS}
\tablehead{
\colhead{Transition} & \colhead{$\lambda_{\rm rest}$ ($\AA$)} &
\colhead{$f$}}

\startdata
NiII 1370 & 1370.132 & 0.1309 \nl
SiIV 1393 & 1393.755 & 0.528 \nl
SnII 1400 & 1400.400 & 0.71 \nl
SiIV 1402 & 1402.770 & 0.262 \nl
GaII 1414 & 1414.402 & 1.8 \nl
SiII 1526 & 1526.707 & 0.2303 \nl
CIV 1548 & 1548.195 & 0.1908 \nl
CIV 1550 & 1550.770 & 0.09522 \nl
GeII 1602 & 1602.4863 & 0.135 \nl
FeII 1608 & 1608.4449 & 0.05545 \nl
AlII 1670 & 1670.7874 & 1.88 \nl
PbII 1682 & 1682.15 & 0.156 \nl
NiII 1741 & 1741.549 & 0.1035 \nl
NiII 1751 & 1751.910 & 0.0638 \nl
SiII 1808 & 1808.0126 & 0.00218 \nl
AlIII 1854 & 1854.716 & 0.539 \nl
AlIII 1862 & 1862.790 & 0.268 \nl
ZnII 2026 & 2026.136 & 0.515 \nl
CrII 2056 & 2056.254 & 0.1403 \nl
CrII 2062 & 2062.234 & 0.1049 \nl
ZnII 2062 & 2062.664 & 0.2529 \nl
FeII 2344 &  2344.214 & 0.1097 \nl
FeII 2374 & 2374.4612 & 0.02818 \nl
FeII 2382 & 2382.765 &  0.3006 \nl
FeII 2600 & 2600.1729 & 0.2239 \nl
MgII 2796 & 2796.352 & 0.6123 \nl
MgII 2803 & 2803.531 & 0.3054 \nl
\enddata
\end{deluxetable}

\begin{deluxetable}{ccccclcc}
\tablewidth{0pc}
\tablenum{4a} 
\tablecaption{FIT FOR $z$=2.462 -- LOW IONS}
\tablehead{
\colhead{Comp} & \colhead{$z$} &
\colhead{$\sigma_z$} & \colhead{$b$} &
\colhead{$\sigma_b$} & \colhead{ION} &
\colhead{log $N$} & \colhead{$\sigma_{\log N}$} \\
& & \colhead{$\sci{-5}$} & \colhead{(\kms)} &
\colhead{(\kms)} & & \colhead{($\cm2$)} & \colhead{($\cm2$)}}

\tiny
\startdata
 1 & 2.460470 &  2.0 &  4.59 &  2.19 & Fe$^+$   & 13.11 &  0.15 \nl
 2 & 2.460598 &  0.3 &  4.44 &  0.47 & Fe$^+$   & 14.11 &  0.05 \nl
 & & & & & Si$^+$   & 14.41 &  0.04 \nl
 & & & & & Ni$^+$   & 12.56 &  0.04 \nl
 & & & & & Cr$^+$   & 12.02 &  0.10 \nl
 3 & 2.460807 &  0.5 &  5.91 &  1.01 & Fe$^+$   & 13.99 &  0.13 \nl
 & & & & & Si$^+$   & 14.36 &  0.16 \nl
 & & & & & Ni$^+$   & 12.30 &  0.32 \nl
 & & & & & Cr$^+$   & 12.18 &  0.12 \nl
 4 & 2.460944 &  5.9 & 12.47 &  4.97 & Fe$^+$   & 13.80 &  0.24 \nl
 & & & & & Si$^+$   & 14.21 &  0.25 \nl
 & & & & & Ni$^+$   & 12.54 &  0.22 \nl
 & & & & & Cr$^+$   & 11.61 &  0.55 \nl
 5 & 2.461347 &  1.4 & 13.94 &  1.17 & Fe$^+$   & 13.99 &  0.04 \nl
 & & & & & Si$^+$   & 14.52 &  0.06 \nl
 & & & & & Ni$^+$   & 12.64 &  0.07 \nl
 & & & & & Cr$^+$   & 12.28 &  0.12 \nl
 6 & 2.461426 &  0.5 &  3.00 &  1.22 & Fe$^+$   & 13.74 &  0.10 \nl
 & & & & & Si$^+$   & 13.94 &  0.17 \nl
 & & & & & Ni$^+$   & 12.00 &  0.19 \nl
 & & & & & Cr$^+$   & 11.69 &  0.31 \nl
 7 & 2.461735 &  0.4 &  8.56 &  0.63 & Fe$^+$   & 14.20 &  0.02 \nl
 & & & & & Si$^+$   & 14.76 &  0.03 \nl
 & & & & & Ni$^+$   & 12.85 &  0.03 \nl
 & & & & & Cr$^+$   & 12.23 &  0.08 \nl
 8 & 2.461905 &  0.4 &  3.24 &  0.89 & Fe$^+$   & 13.92 &  0.08 \nl
 & & & & & Si$^+$   & 14.33 &  0.07 \nl
 & & & & & Ni$^+$   & 12.45 &  0.07 \nl
 & & & & & Cr$^+$   & 11.96 &  0.13 \nl
 9 & 2.462091 &  1.4 &  8.23 &  2.59 & Fe$^+$   & 13.68 &  0.11 \nl
 & & & & & Si$^+$   & 14.19 &  0.14 \nl
 & & & & & Ni$^+$   & 12.30 &  0.12 \nl
 & & & & & Cr$^+$   & 11.06 &  1.15 \nl
10 & 2.462266 &  1.5 &  6.22 &  2.42 & Fe$^+$   & 13.51 &  0.19 \nl
 & & & & & Si$^+$   & 14.13 &  0.19 \nl
 & & & & & Ni$^+$   & 11.94 &  0.30 \nl
 & & & & & Cr$^+$   & 11.02 &  1.22 \nl
11 & 2.462494 &  2.2 & 11.31 &  2.43 & Fe$^+$   & 13.99 &  0.10 \nl
 & & & & & Si$^+$   & 14.59 &  0.11 \nl
 & & & & & Ni$^+$   & 12.59 &  0.11 \nl
 & & & & & Cr$^+$   & 12.27 &  0.10 \nl
12 & 2.462594 &  0.6 &  3.00 &  1.49 & Fe$^+$   & 13.88 &  0.18 \nl
 & & & & & Si$^+$   & 14.17 &  0.17 \nl
 & & & & & Ni$^+$   & 12.18 &  0.18 \nl
13 & 2.462818 &  0.6 & 14.58 &  0.56 & Fe$^+$   & 14.43 &  0.02 \nl
 & & & & & Si$^+$   & 14.92 &  0.02 \nl
 & & & & & Ni$^+$   & 13.00 &  0.02 \nl
 & & & & & Cr$^+$   & 12.57 &  0.05 \nl
\enddata
\normalsize
\end{deluxetable}

\begin{table}
\dummytable\tablenum{2}\label{orders}
\end{table}

\begin{table}
\dummytable\tablenum{3}\label{fval}
\end{table}

\begin{table}
\dummytable\tablenum{4a}\label{2462TL}
\end{table}

\begin{table*}
\begin{center}
\begin{tabular}{ccccclcc}
Comp & $z$ & $\sigma_z$ & b & $\sigma_b$ & Ion & log $N$ &
$\sigma_{\rm{log} {\it N}}$\cr
&
& ($\sci{-5}$)
& (\kms)
& (\kms)
&
& ($\cm2$)
& ($\cm2$) \cr
\tableline
 1 & 2.459346 &  2.5 & 18.33 &  3.03 & Si$^{+3}$ & 11.72 &  0.53 \cr
 & & & & & C$^{+3}$  & 13.47 &  0.07 \cr
 2 & 2.459713 &  4.4 & 43.14 &  8.77 & Si$^{+3}$ & 12.81 &  0.12 \cr
 & & & & & C$^{+3}$  & 13.95 &  0.04 \cr
 3 & 2.460433 &  1.0 &  9.43 &  0.99 & Si$^{+3}$ & 12.77 &  0.07 \cr
 & & & & & C$^{+3}$  & 13.45 &  0.10 \cr
 4 & 2.460753 &  1.2 & 16.65 &  1.50 & Si$^{+3}$ & 13.45 &  0.04 \cr
 & & & & & C$^{+3}$  & 14.19 &  0.13 \cr
 5 & 2.461035 &  0.8 & 11.63 &  0.64 & Si$^{+3}$ & 13.54 &  0.04 \cr
 & & & & & C$^{+3}$  & 13.10 &  0.24 \cr
 6 & 2.461647 & 22.3 & 82.54 & 22.64 & Si$^{+3}$ & 13.36 &  0.14 \cr
 & & & & & C$^{+3}$  & 12.85 &  0.38 \cr
 7 & 2.461863 &  0.9 & 15.01 &  1.61 & Si$^{+3}$ & 12.81 &  0.08 \cr
 & & & & & C$^{+3}$  & 13.41 &  0.48 \cr
 8 & 2.462229 &  3.5 & 11.01 &  5.97 & Si$^{+3}$ & 11.67 &  0.52 \cr
 9 & 2.462762 &  9.1 & 38.26 & 11.23 & Si$^{+3}$ & 12.81 &  0.31 \cr
 & & & & & C$^{+3}$  & 13.35 &  0.11 \cr
10 & 2.462908 &  2.1 &  5.38 &  5.93 & Si$^{+3}$ & 11.59 &  0.34 \cr 
& & & & & C$^{+3}$  & 13.63 &  0.12 \cr

\end{tabular}
\end{center}

\tablenum{4b}
\caption{FIT FOR $z$=2.462 -- HIGH IONS} \label{2462H}
\end{table*}

\clearpage

\begin{planotable}{ccccclcc}
\tablewidth{350.0pt}
\tablenum{5a} \label{tbl-4a}
\tablecaption{FIT FOR $z$=2.457 -- LOW IONS}
\tablehead{
\colhead{Comp} & \colhead{$z$} &
\colhead{$\sigma_z$} & \colhead{$b$} &
\colhead{$\sigma_b$} & \colhead{ION} &
\colhead{log $N$} & \colhead{$\sigma_{\log N}$} \nl
& & \colhead{$\sci{-5}$} & \colhead{(\kms)} &
\colhead{(\kms)} & & \colhead{($\cm2$)} & \colhead{($\cm2$)}}
\small
\startdata
1 & 2.456041 &  0.5 &  3.95 &  0.80 & Si$^{+}$   & 12.53 &  0.06 \nl
 & & & & & Fe$^+$     & 12.34 &  0.20 \nl
 & & & & & Al$^+$     & 11.81 &  0.04 \nl
 2 & 2.456295 &  0.3 &  9.58 &  0.49 & Si$^{+}$   & 13.40 &  0.02 \nl
 & & & & & Fe$^+$     & 13.19 &  0.04 \nl
 & & & & & Al$^+$     & 12.60 &  0.02 \nl
 3 & 2.456540 &  1.5 &  5.44 &  2.57 & Si$^{+}$   & 12.54 &  0.15 \nl
 & & & & & Fe$^+$     & 12.30 &  0.29 \nl
 & & & & & Al$^+$     & 11.78 &  0.12 \nl
 4 & 2.456679 &  0.6 &  3.30 &  1.30 & Si$^{+}$   & 13.03 &  0.05 \nl
 & & & & & Fe$^+$     & 12.78 &  0.12 \nl
 & & & & & Al$^+$     & 12.19 &  0.06 \nl
 5 & 2.456844 &  1.6 &  8.95 &  2.88 & Si$^{+}$   & 12.98 &  0.18 \nl
 & & & & & Fe$^+$     & 12.95 &  0.14 \nl
 & & & & & Al$^+$     & 12.26 &  0.16 \nl
 6 & 2.457079 &  1.5 & 12.83 &  1.78 & Si$^{+}$   & 13.41 &  0.06 \nl
 & & & & & Fe$^+$     & 13.04 &  0.10 \nl
 & & & & & Al$^+$     & 12.65 &  0.06 \nl
 7 & 2.457336 &  0.6 &  3.00 &  1.15 & Si$^{+}$   & 13.21 &  0.11 \nl
 & & & & & Fe$^+$     & 12.78 &  0.09 \nl
 & & & & & Al$^+$     & 12.28 &  0.11 \nl
 8 & 2.457460 &  0.6 &  3.00 &  1.22 & Si$^{+}$   & 13.16 &  0.10 \nl
 & & & & & Fe$^+$     & 12.63 &  0.12 \nl
 & & & & & Al$^+$     & 12.22 &  0.09 \nl
 9 & 2.457689 &  6.4 & 25.16 &  7.00 & Si$^{+}$   & 13.04 &  0.13 \nl
 & & & & & Al$^+$     & 12.37 &  0.11 \nl
10 & 2.457858 &  0.3 &  6.37 &  0.57 & Si$^{+}$   & 13.35 &  0.04 \nl
 & & & & & Fe$^+$     & 13.11 &  0.05 \nl
 & & & & & Al$^+$     & 12.42 &  0.05 \nl
11 & 2.458100 &  1.1 &  4.65 &  2.29 & Si$^{+}$   & 12.40 &  0.15 \nl
 & & & & & Fe$^+$     & 12.17 &  0.30 \nl
 & & & & & Al$^+$     & 11.42 &  0.25 \nl
12 & 2.458406 &  0.8 & 14.96 &  1.20 & Si$^{+}$   & 12.80 &  0.05 \nl
 & & & & & Al$^+$     & 12.09 &  0.03 \nl
\enddata
\normalsize
\end{planotable}

\begin{table}
\dummytable\tablenum{5a}\label{2457L}
\end{table}

\clearpage

\begin{table*}
\begin{center}
\begin{tabular}{ccccclcc}
Comp & $z$ & $\sigma_z$ & b & $\sigma_b$ & Ion & log $N$ &
$\sigma_{\rm{log} {\it N}}$\cr
&
& ($\sci{-5}$)
& (\kms)
& (\kms)
&
& ($\cm2$)
& ($\cm2$) \cr
\tableline
 1 & 2.456343 &  0.7 & 24.36 &  0.78 & Si$^{+3}$ & 12.98 &  0.02 \cr
 & & & & & C$^{+3}$  & 13.86 &  0.02 \cr
 2 & 2.457273 &  0.5 & 34.58 &  0.70 & Si$^{+3}$ & 13.65 &  0.01 \cr
 & & & & & C$^{+3}$  & 14.54 &  0.02 \cr
 3 & 2.457873 &  0.4 &  8.77 &  0.61 & Si$^{+3}$ & 12.98 &  0.02 \cr
 & & & & & C$^{+3}$  & 13.75 &  0.05 \cr
 4 & 2.458106 &  0.9 &  7.53 &  1.17 & Si$^{+3}$ & 12.45 &  0.06 \cr
 & & & & & C$^{+3}$  & 13.32 &  0.07 \cr
 5 & 2.458454 &  1.5 & 30.67 &  1.98 & Si$^{+3}$ & 12.67 &  0.04 \cr
 & & & & & C$^{+3}$  & 13.77 &  0.03 \cr
\end{tabular}
\end{center}

\tablenum{5b}
\caption{FIT FOR $z$=2.457 -- HIGH IONS} \label{2457H}
\end{table*}

\clearpage

\begin{planotable}{ccccclcc}
\tablewidth{350.0pt}
\tablenum{6a} \label{tbl-6a}
\tablecaption{FIT FOR $z$=2.325 -- LOW IONS} 
\tablehead{
\colhead{Comp} & \colhead{$z$} &
\colhead{$\sigma_z$} & \colhead{$b$} &
\colhead{$\sigma_b$} & \colhead{ION} &
\colhead{log $N$} & \colhead{$\sigma_{\log N}$} \nl
& & \colhead{$\sci{-5}$} & \colhead{(\kms)} &
\colhead{(\kms)} & & \colhead{($\cm2$)} & \colhead{($\cm2$)}}
\tiny
\startdata
 1 & 2.323701 &  0.6 &  6.87 &  0.99 & Al$^+$     & 11.52 &  0.06 \nl
 & & & & & Al$^{+2}$  & 11.97 &  0.05 \nl
 & & & & & Fe$^+$     & 11.87 &  0.78 \nl
 & & & & & Si$^+$     & 12.73 &  0.04 \nl
 2 & 2.323896 &  0.3 &  3.39 &  0.57 & Al$^+$     & 13.00 &  0.25 \nl
 & & & & & Al$^{+2}$  & 12.59 &  0.03 \nl
 & & & & & Fe$^+$     & 13.32 &  0.04 \nl
 & & & & & Si$^+$     & 13.92 &  0.24 \nl
 3 & 2.324033 &  0.3 &  3.37 &  0.69 & Al$^+$     & 12.83 &  0.23 \nl
 & & & & & Al$^{+2}$  & 12.48 &  0.03 \nl
 & & & & & Fe$^+$     & 13.30 &  0.04 \nl
 & & & & & Si$^+$     & 13.81 &  0.26 \nl
 4 & 2.324173 &  1.0 &  3.41 &  1.86 & Al$^{+2}$  & 11.81 &  0.07 \nl
 5 & 2.324195 &  2.8 &  9.92 &  3.16 & Al$^+$     & 11.63 &  0.11 \nl
 & & & & & Si$^+$     & 12.22 &  0.17 \nl
 6 & 2.324663 &  0.2 &  6.24 &  0.26 & Al$^+$     & 12.28 &  0.02 \nl
 & & & & & Al$^{+2}$  & 12.02 &  0.04 \nl
 & & & & & Fe$^+$     & 12.71 &  0.13 \nl
 & & & & & Si$^+$     & 13.21 &  0.02 \nl
 7 & 2.324884 &  1.2 &  2.80 &  2.97 & Al$^+$     & 11.11 &  0.12 \nl
 & & & & & Si$^+$     & 11.98 &  0.15 \nl
 8 & 2.325078 &  0.7 &  3.12 &  0.88 & Al$^+$     & 11.97 &  0.05 \nl
 & & & & & Al$^{+2}$  & 11.92 &  0.06 \nl
 & & & & & Fe$^+$     & 12.47 &  0.22 \nl
 & & & & & Si$^+$     & 13.06 &  0.06 \nl
 9 & 2.325180 &  0.4 &  2.27 &  0.78 & Al$^+$     & 13.03 &  0.70 \nl
 & & & & & Al$^{+2}$  & 11.99 &  0.06 \nl
 & & & & & Fe$^+$     & 13.36 &  0.07 \nl
 & & & & & Si$^+$     & 14.30 &  0.76 \nl
10 & 2.325560 &  0.6 &  5.56 &  1.02 & Al$^+$     & 11.45 &  0.06 \nl
 & & & & & Al$^{+2}$  & 11.65 &  0.09 \nl
 & & & & & Fe$^+$     & 12.01 &  0.56 \nl
 & & & & & Si$^+$     & 12.47 &  0.06 \nl
11 & 2.325796 &  1.5 &  4.88 &  2.55 & Al$^+$     & 11.20 &  0.10 \nl
12 & 2.326045 &  0.3 &  4.68 &  0.39 & Al$^+$     & 12.16 &  0.03 \nl
 & & & & & Al$^{+2}$  & 12.11 &  0.04 \nl
 & & & & & Fe$^+$     & 12.46 &  0.20 \nl
 & & & & & Si$^+$     & 13.34 &  0.03 \nl
13 & 2.326198 &  0.5 &  6.88 &  0.54 & Al$^+$     & 12.08 &  0.03 \nl
 & & & & & Al$^{+2}$  & 12.09 &  0.04 \nl
 & & & & & Fe$^+$     & 12.75 &  0.12 \nl
 & & & & & Si$^+$     & 13.21 &  0.03 \nl
\enddata
\normalsize
\end{planotable}

\begin{table}
\dummytable\tablenum{6a}\label{2325TL}
\end{table}

\clearpage

\begin{table*}
\begin{center}
\begin{tabular}{ccccclcc}
Comp & $z$ & $\sigma_z$ & b & $\sigma_b$ & Ion & log $N$ &
$\sigma_{\rm{log} {\it N}}$\cr
&
& ($\sci{-5}$)
& (\kms)
& (\kms)
&
& ($\cm2$)
& ($\cm2$) \cr
\tableline
 1 & 2.319744 &  0.2 &  9.11 &  0.29 & C$^{+3}$  & 13.64 &  0.01 \cr
 2 & 2.323508 &  4.4 & 15.48 &  3.55 & C$^{+3}$  & 13.13 &  0.13 \cr
 3 & 2.323736 &  1.2 &  9.13 &  1.83 & C$^{+3}$  & 13.47 &  0.16 \cr
 4 & 2.324048 &  1.4 & 24.68 &  2.38 & C$^{+3}$  & 14.25 &  0.04 \cr
 5 & 2.324569 &  2.4 & 15.07 &  3.74 & C$^{+3}$  & 13.56 &  0.13 \cr
 6 & 2.324825 &  1.9 & 11.15 &  2.60 & C$^{+3}$  & 13.41 &  0.18 \cr
 7 & 2.325100 &  3.8 & 16.40 &  4.30 & C$^{+3}$  & 13.25 &  0.12 \cr
 8 & 2.325624 &  0.7 & 18.12 &  1.23 & C$^{+3}$  & 13.47 &  0.02 \cr
 9 & 2.326123 &  0.2 & 14.21 &  0.25 & C$^{+3}$  & 14.23 &  0.01 \cr
\end{tabular}
\end{center}

\tablenum{6b}
\caption{ FIT FOR $z$=2.325 -- HIGH IONS} \label{2325H}
\end{table*}

\begin{table*}
\begin{center}
\begin{tabular}{ccccclcc}
Comp & $z$ & $\sigma_z$ & b & $\sigma_b$ & Ion & log $N$ &
$\sigma_{\rm{log} {\it N}}$\cr
&
& ($\sci{-5}$)
& (\kms)
& (\kms)
&
& ($\cm2$)
& ($\cm2$) \cr
\tableline
 1 & 1.475760 &  1.5 &  1.21 & 5.88 & Fe$^+$ & 12.24 & 0.18 \cr
 2 & 1.475858 &  0.7 &  4.27 & 1.87 & Fe$^+$ & 12.71 & 0.07 \cr
 3 & 1.476039 &  0.4 &  3.59 & 0.52 & Fe$^+$ & 13.65 & 0.03 \cr
 4 & 1.476115 &  0.8 &  2.44 & 1.37 & Fe$^+$ & 13.22 & 0.06 \cr
 5 & 1.476182 &  0.9 &  2.05 & 1.86 & Fe$^+$ & 12.92 & 0.08 \cr
 6 & 1.476301 &  0.1 &  2.70 & 0.22 & Fe$^+$ & 13.64 & 0.04 \cr
\end{tabular}
\end{center}

\tablenum{7}
\caption{FIT FOR $z$=1.476} \label{1476}
\end{table*}

\begin{table*}
\begin{center}
\begin{tabular}{lccc}
Transition
& Apparent
& VPFIT 
& Adopted \cr
\tableline
Si IV 1393    & $14.039  \pm 0.006$ & $14.050 \pm 0.038$ & $14.045 \pm 0.038$ \cr
C IV 1550    & $14.617  \pm 0.005$ & $14.615 \pm 0.062$ & $ 14.616 \pm 0.062$ \cr
\cr
C I 1560     & $13.228  \pm 0.059$ & & $13.228  \pm 0.059$ \cr
Fe II 1608   & $15.033  \pm 0.004$ & $15.086 \pm 0.024$ & $15.060 \pm 0.024$ \cr
Al II 1671    \cr
Ni II 1709    & $13.785  \pm 0.020$ & & $13.785  \pm 0.020$ \cr
Ni II 1741    & $13.638  \pm 0.013$ & $13.631 \pm 0.030$ & $ 13.628
\pm 0.030 $ \cr
Ni II 1751    & $13.610  \pm 0.020$ & & \cr
Si II 1808    & $15.561  \pm 0.010$ & $15.546 \pm 0.026$ & $15.554 \pm 0.026$ \cr
Al III 1862   & $13.606  \pm 0.008$ & & $13.606  \pm 0.008$ \cr
Cr II 2056    & $13.158  \pm 0.030$ & $13.138 \pm 0.040$ & $13.148 \pm 0.040$ \cr
Cr II 2062    & $13.401  \pm 0.025$ & & $13.401  \pm 0.025$ \cr
Zn II 2062    & $12.468  \pm 0.046$ & & $12.468  \pm 0.046$ \cr
\cr
Pb II 1682   &  $12.66 \pm 0.114$ & & $12.66 \pm 0.114$ \cr
Ge II 1602   &  $< 12.65$ & & $< 12.65$ \cr
\end{tabular}
\end{center}

\tablenum{8}
\caption{IONIC COLUMN DENSITIES FOR $z$ = 2.462} \label{2462I}

\tablecomments{Values reported in logarithmic space have 
deceptively small errors.  For
instance, the value for $\N{Pb}$ is not a $5 \sigma$ detection and that for
$\N{Ge}$ is not even a $1 \sigma$ detection.}

\end{table*}

\begin{table*}
\begin{center}
\begin{tabular}{lccc}
Transition
& Apparent
& VPFIT
& Adopted \cr
\tableline
Si IV 1393    & $13.839  \pm 0.003$ & $13.85 \pm 0.01$ & $ 13.85 \pm 0.01$ \cr
Si IV 1403    & $13.851  \pm 0.007$ & & $13.85 \pm 0.01$ \cr
C IV 1550    & $14.686  \pm 0.041$ & $14.74 \pm 0.01$ & $ 14.737
\pm 0.010 $ \cr
\cr
Si II 1526    & $14.129  \pm 0.006$ & $14.18 \pm 0.02$ & $ 14.18 \pm 0.02$ \cr
Fe II 1608    & $13.839  \pm 0.022$ & $13.85 \pm 0.03$ & $ 13.843
\pm 0.018 $ \cr
Al II 1671    & $13.321  \pm 0.004$ & $13.36 \pm 0.02$ & $ 13.36 \pm 0.02 $ \cr
Al III 1862   & $12.920  \pm 0.028$ & & $12.920  \pm 0.028$ \cr
\end{tabular}
\end{center}

\tablenum{9}
\caption{IONIC COLUMN DENSITIES FOR $z$ = 2.457} \label{2457I}
\end{table*}

\begin{table*}
\begin{center}
\begin{tabular}{cccl}
Component 
& log$_{10} \> N$(HI)
& $z_{\rm abs}$
& $b$ (\kms) \cr
\tableline
           1 &   18.38 & 2.460470 &  4.59 \cr
           2 &   19.38 & 2.460598 &  4.44 \cr
           3 &   19.26 & 2.460807 &  5.91 \cr
           4 &   19.07 & 2.460944 & 12.47 \cr
           5 &   19.26 & 2.461347 & 13.94 \cr
           6 &   19.01 & 2.461426 &  3.00 \cr
           7 &   19.47 & 2.461735 &  8.56 \cr
           8 &   19.18 & 2.461905 &  3.24 \cr
           9 &   18.95 & 2.462091 &  8.23 \cr
          10 &   18.78 & 2.462266 &  6.22 \cr
          11 &   19.26 & 2.462494 & 11.31 \cr
          12 &   19.15 & 2.462594 &  3.00 \cr
          13 &   19.70 & 2.462818 & 14.58 \cr
\cr
          14 &   17.61 & 2.456041 &  3.95 \cr
          15 &   18.46 & 2.456295 &  9.58 \cr
          16 &   17.57 & 2.456540 &  5.44 \cr
          17 &   18.05 & 2.456679 &  3.30 \cr
          18 &   18.22 & 2.456844 &  8.95 \cr
          19 &   18.31 & 2.457079 & 12.83 \cr
          20 &   18.05 & 2.457336 &  3.00 \cr
          21 &   17.90 & 2.457460 &  3.00 \cr
          22 &   18.38 & 2.457858 &  6.37 \cr
          23 &   17.44 & 2.458100 &  4.65 \cr
\end{tabular}
\end{center}

\tablenum{10}
\caption{HI COMPONENTS IN \Lya PROFILE ($z$=2.46)} \label{HI}
\end{table*}

\begin{table*}
\begin{center}
\begin{tabular}{cccccc}
Velocity (\kms) & [Si$^+$/Si$^{3+}$] & [H$^0$/H$^+$]$_{\rm Si}$
& [Al$^+$/Al$^{++}$] & [H$^0$/H$^+$]$_{\rm Al}$ &
[H$^0$/H$^+$]$_{\rm Al \; + \; Si}$ \cr
\tableline
$-218 < v < -140$ & 1.0 & $-0.16$ & 0.42 & $-0.06$ & $-0.06$ \cr
$-140 < v < -110$ & 1.7 & $0.12$ & 0.47 & $0.00$ & $0.12$ \cr
$-110 < v < -40$ & 1.8 &  $0.13$ & 0.43 & $-0.04$ & $0.13$ \cr
$-40 < v < 40$   & 2.1 &  $0.31$ & 0.27 & $-0.30$ & $0.31$ \cr
\end{tabular}
\end{center}

\tablenum{11}
\caption{IONIZATION LIMITS} \label{rtio_tab}

\tablecomments{All values are conservative lower limits}
\end{table*}

\begin{table*}
\begin{center}
\begin{tabular}{lccc}
Metal
& log$_{10} N$(X) (cm$^-2$)
& [X/H] \hfil \cr
\tableline
Fe & $15.060 \pm 0.024$ & $-0.830 \pm 0.051$ \cr
Ni & $13.628 \pm 0.030$ & $-1.002 \pm 0.054$ \cr
Al & SATU \cr
Si & $15.554 \pm 0.026$ & $-0.376 \pm  0.052$ \cr
Cr & $13.158 \pm 0.030$ & $-0.902 \pm 0.064$ \cr
Zn & $> 12.468 \pm 0.046$ & $> -0.562 \pm 0.064$ \cr
\cr
C  & $14.616 \pm 0.062$ & $-2.324 \pm 0.077$ \cr
Si & $14.045 \pm 0.038$ & $-1.885 \pm 0.059$ \cr
\cr
Pb &  $12.66 \pm 0.114$ & $2.23 \pm 0.121$ \cr
Ge  &  $< 12.65$ & $< 0.644$ \cr
\end{tabular}
\end{center}

\tablenum{12}
\caption{ABUNDANCES FOR $z$ = 2.462} \label{abnd}
\end{table*}

\begin{table*}
\begin{center}
\begin{tabular}{cccc}
[X/Zn]
& Feature 1
& Feature 2
& Feature 3 \cr
& $-144 \leftrightarrow -110$ km/s
& $-40 \leftrightarrow -9$ km/s
& $-9 \leftrightarrow 35$ km/s \cr
\tableline
Fe & $-0.856 \pm 0.185$ & $-0.679 \pm 0.175$ & $-0.508 \pm 0.243$ \cr
Ni & $-0.939 \pm 0.204$ & $-0.831 \pm 0.186$ & $-0.638 \pm 0.249$ \cr
Si & $-0.388 \pm 0.197$ & $-0.157 \pm 0.180$ & $-0.035 \pm 0.246$ \cr
Cr & $-0.833 \pm 0.261$ & $-0.729 \pm 0.220$ & $-0.534 \pm 0.274$ \cr
\end{tabular}
\end{center}

\tablenum{13}
\caption{DEPLETION RELATIVE TO Zn FOR $z$ = 2.462} \label{depl}
\end{table*}

\begin{table*}
\begin{center}
\begin{tabular}{cccc}
Metal
& $\Delta D$
& $|m|$\tablenotemark{a}
& ${(n_H)_2 \over (n_H)_1}$ \cr
\tableline
Fe & $0.348 \pm 0.305 $ & $ 0.38 \pm 0.05 $ & $ 8.2 \pm 9.7 $ \cr
Si & $0.353 \pm 0.315 $ & $ 0.49 \pm 0.15 $ & $ 5.3 \pm 6.6 $ \cr
Cr & $0.299 \pm 0.378 $ & $ 0.50 \pm 0.11 $ & $ 4.0 \pm 5.3 $ \cr
\end{tabular}
\end{center}

\tablenotetext{a}{$m$ values are taken from Jenkins 1987}

\tablenum{14}
\caption{$n_H$ VARIATIONS FOR $z$=2.462} \label{nH}
\end{table*}

\begin{table*}
\begin{center}
\begin{tabular}{lccccc}
Metal
& Comp 1
& Comp 2
& Comp 3
& Comp 4
& Comp 5 \hfil\cr
& $-211 \leftrightarrow -181$
& $-181 \leftrightarrow -144$
& $-144 \leftrightarrow -110$
& $-110 \leftrightarrow -40 $
& $-40  \leftrightarrow 35$ \hfil\cr
\tableline
Fe & 0.11 & 0.14 & 0.11 & 0.25 & 0.39 \cr
Ni & 0.09 & 0.14 & 0.13 & 0.27 & 0.38 \cr
Si II & 0.10 & 0.11 & 0.11 & 0.28 & 0.42 \cr
Cr & 0.11 & 0.15 & 0.14 & 0.23 & 0.40 \cr
Si IV & 0.17 & 0.46 & 0.09 & 0.14 & 0.10 \cr
\end{tabular}
\end{center}

\tablenum{15}
\caption{PERCENT OF TOTAL ABUNDANCES FOR $z$ = 2.462} \label{per}
\end{table*}

\clearpage

\clearpage

\begin{figure}
\centerline{
\psfig{figure=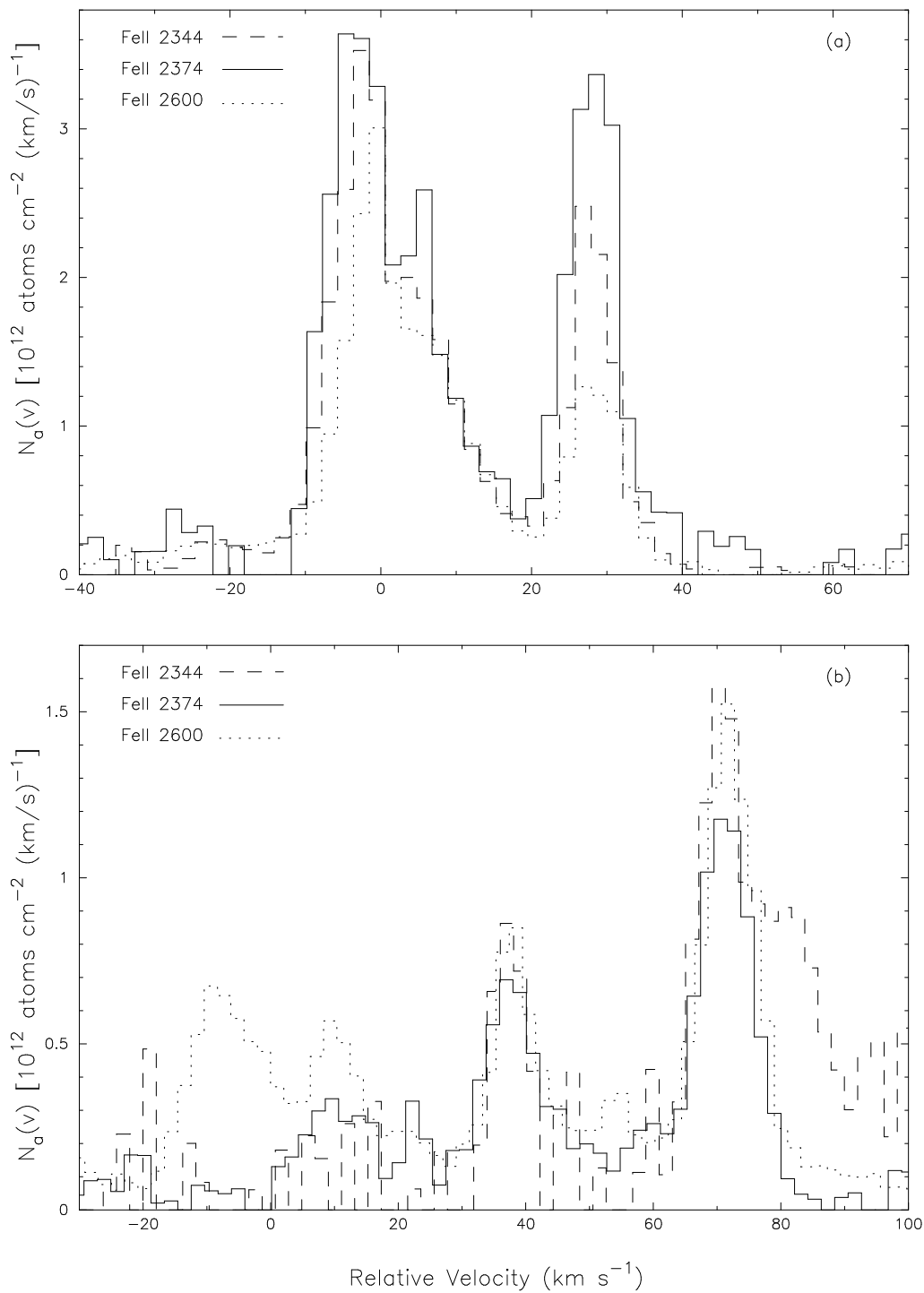,height=7.5in}}
\caption{ Apparent column density $[N_a (v)]$ for Fe II in the
(a) $z$=1.476 and (b) $z$=1.699 systems.  In (a), the regions where Fe II
2374 dominates indicate hidden saturated components.  On the other
hand, the features in (b) where the $N_a (v)$ profiles do not 
coincide are due to
blends with other absorption systems.}
\label{HCA-Fe}
\end{figure}

\begin{figure}
\centerline{
\psfig{figure=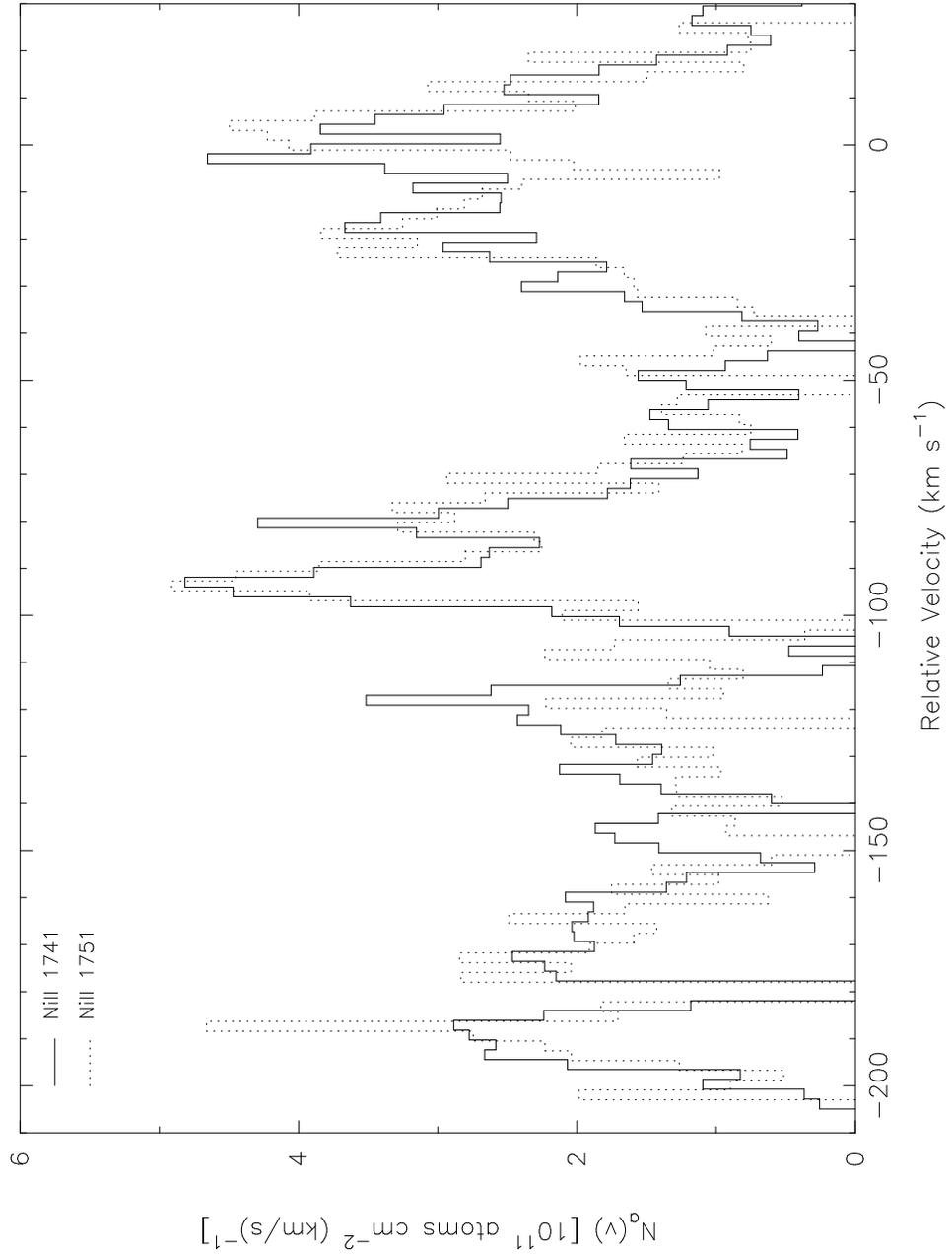,height=7.5in}}
\caption{ Apparent column density $[N_a (v)]$ profiles for Ni II in the $z$=2.462
system.  The good agreement between the two profiles suggests there 
are no hidden saturated components.}
\label{HCA-Ni}
\end{figure}

\begin{figure}
\centerline{
\psfig{figure=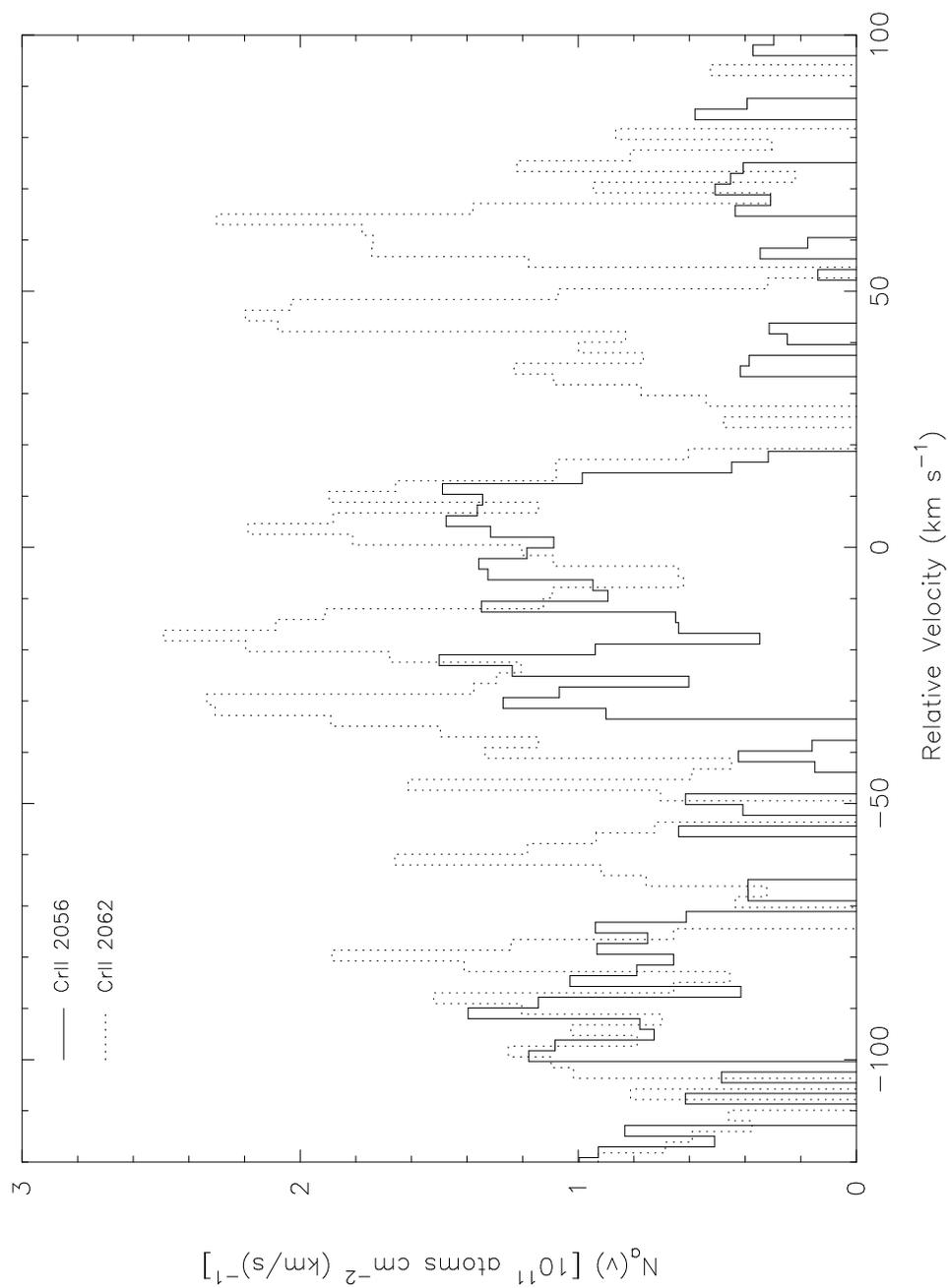,height=7.5in}}
\caption{ Apparent column density $[N_a (v)]$ profiles for Cr II in the $z$=2.462
system.  The Cr II 2062 profile is clearly stronger
over the entire velocity interval.
We suggest this is due to blending in the Cr II 2062 profile by Zn II 2062.
In particular, we believe the features at $-$60, 40 and 60 \kms are due
solely to Zn II absorption.}
\label{HCA-Cr}
\end{figure}

\begin{figure}
\centerline{
\psfig{figure=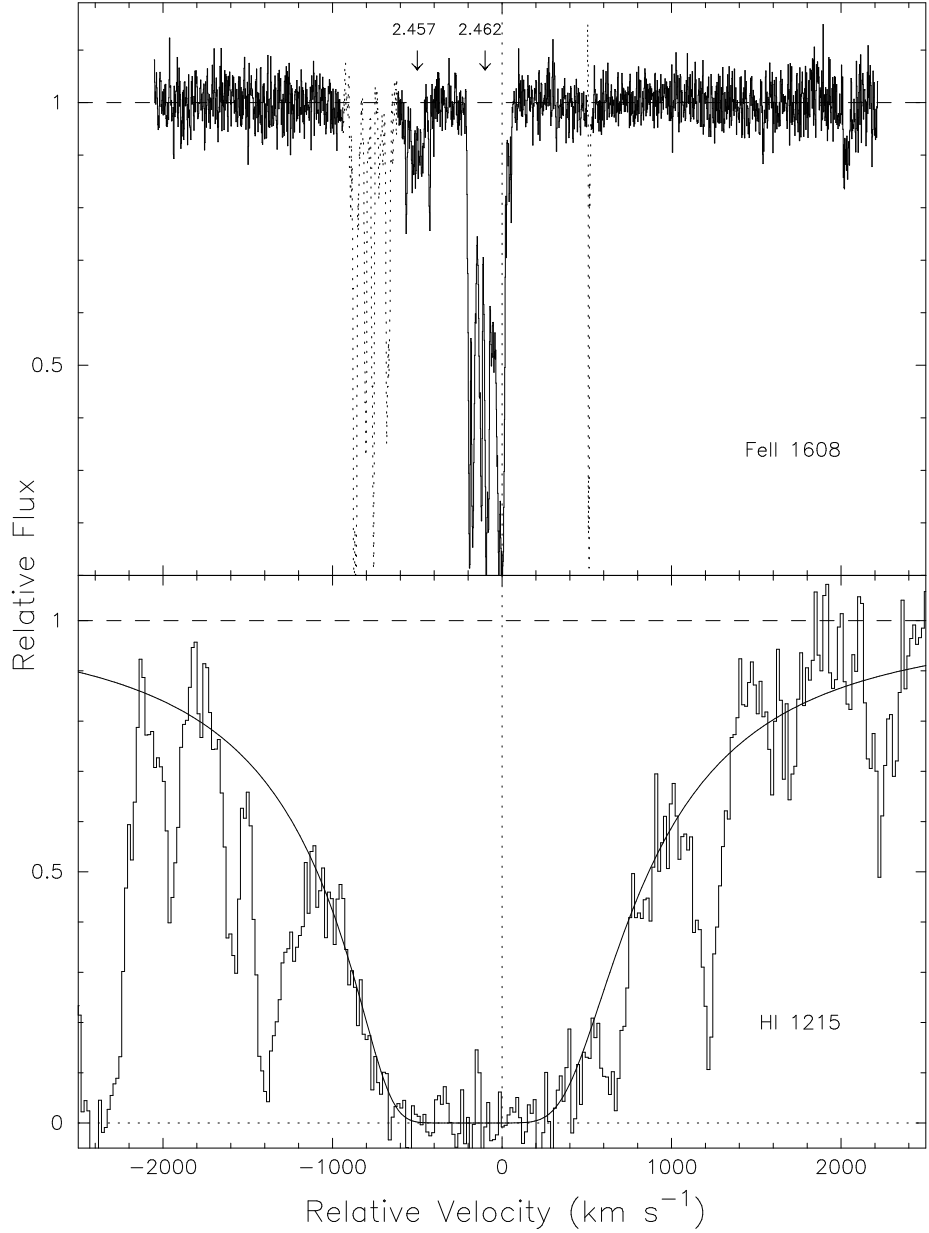,height=7.5in}}
\caption{ Velocity plot of the Fe II 1608 and \Lya transitions 
for the $z$=2.457 and $z$=2.462 systems.
The Voigt profile overplotted on the \Lya velocity plot was derived 
with 23 individual HI
components (See Table 10).}
\label{2462L}
\end{figure}

\begin{figure}
\centerline{
\psfig{figure=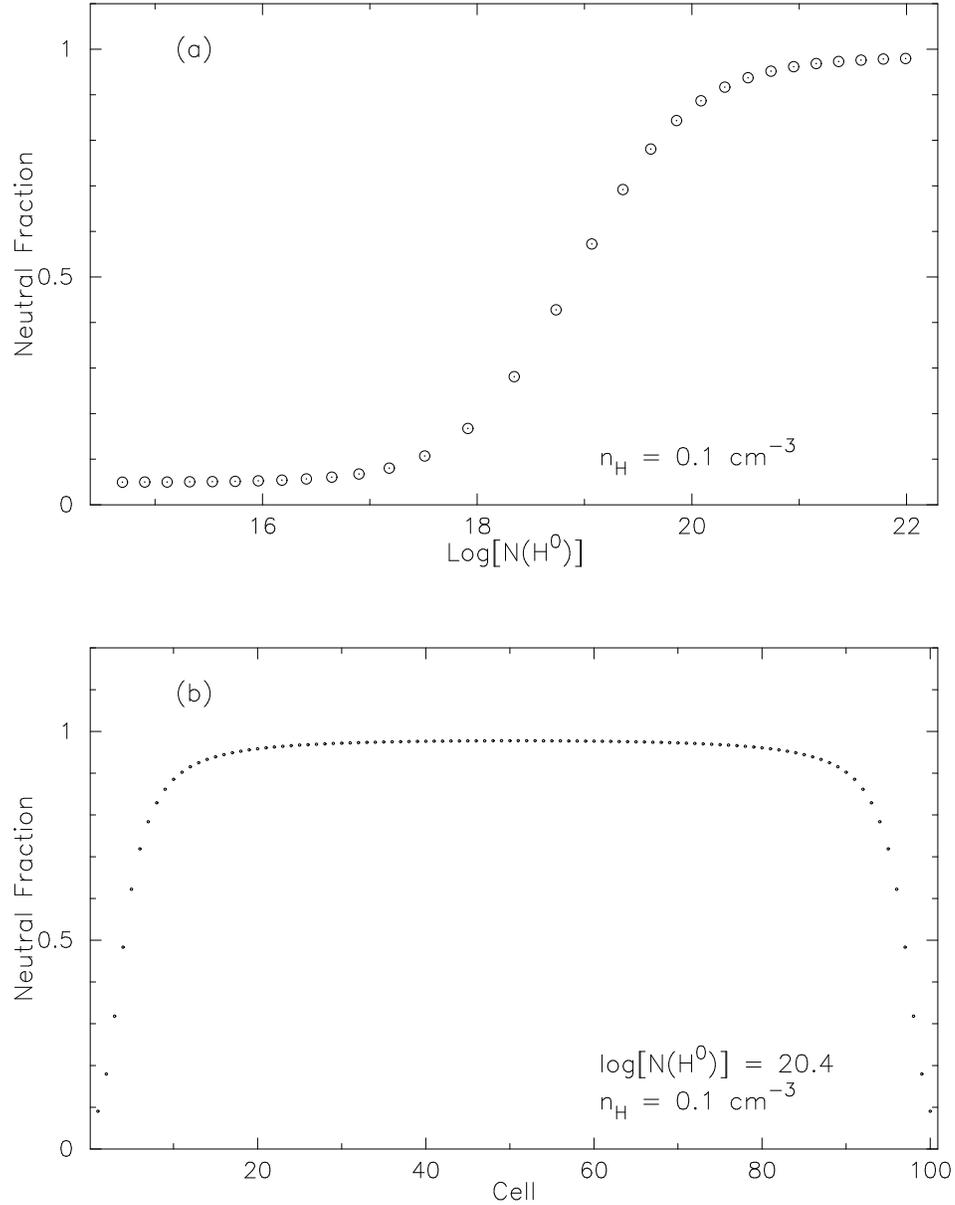,height=7.5in}}
\caption{ (a) Fractional ionization versus log[$\N{HI}$] of a constant
density (n = 0.1 $\rm
cm^{-3}$), plane-parallel Hydrogen disk assuming a temperature of $10^4$ K.
The ionizing spectra is an attenuated power law spectrum, 
calculated by Madau (1992) at
an average redshift of $z$=2.46. (b)
Fractional ionization versus disk depth (expressed in cell number where cell
1 and 100 are the outer faces) for a Hydrogen system with log[$\N{HI}$] = 20.4.  
Note only the edgemost cells are significantly ionized.}
\label{Vince}
\end{figure}

\begin{figure}
\centerline{
\psfig{figure=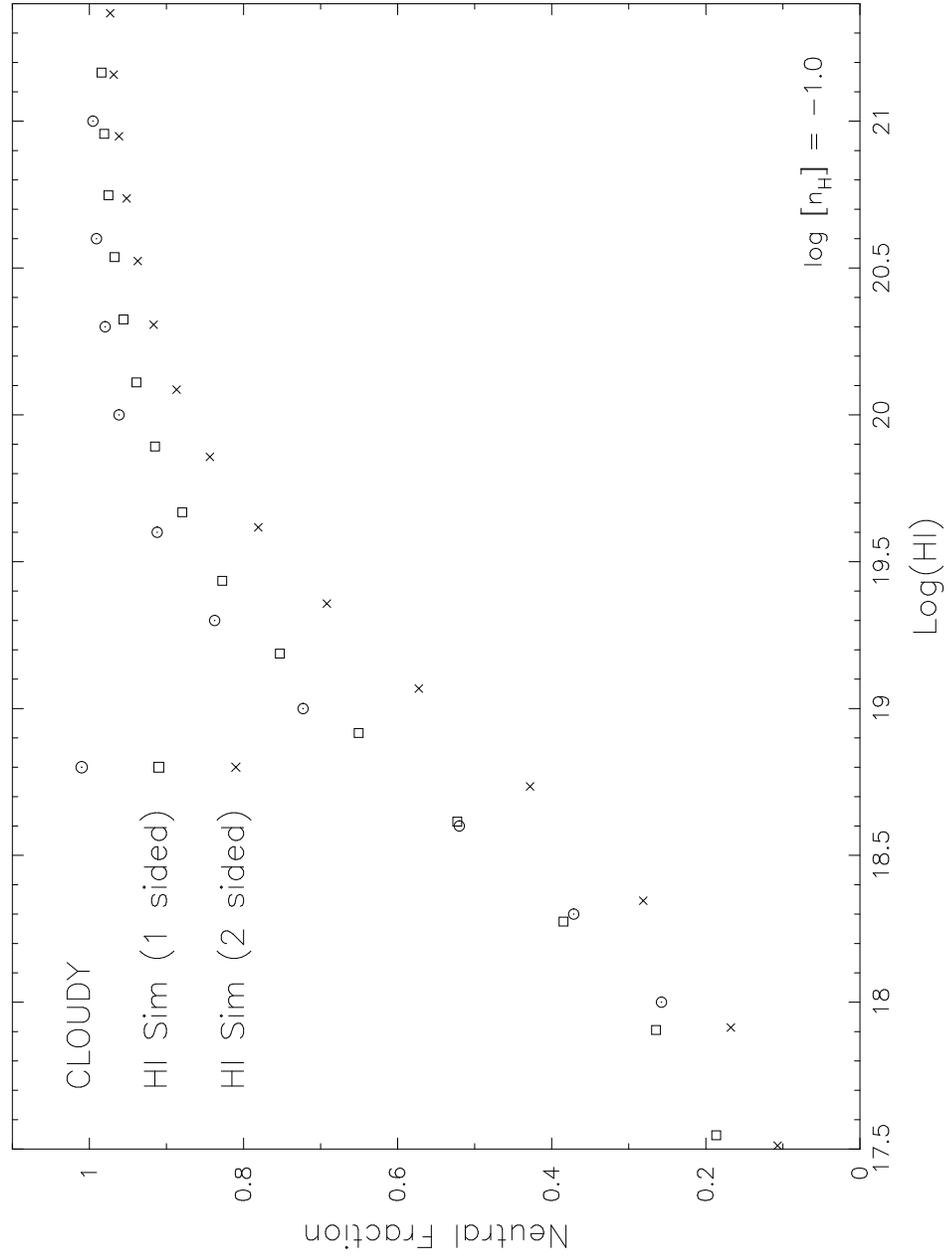,height=7.5in}}
\caption{ Comparison of our ionization model with CLOUDY simulations. }
\label{I-Comp}
\end{figure}

\begin{figure}
\centerline{
\psfig{figure=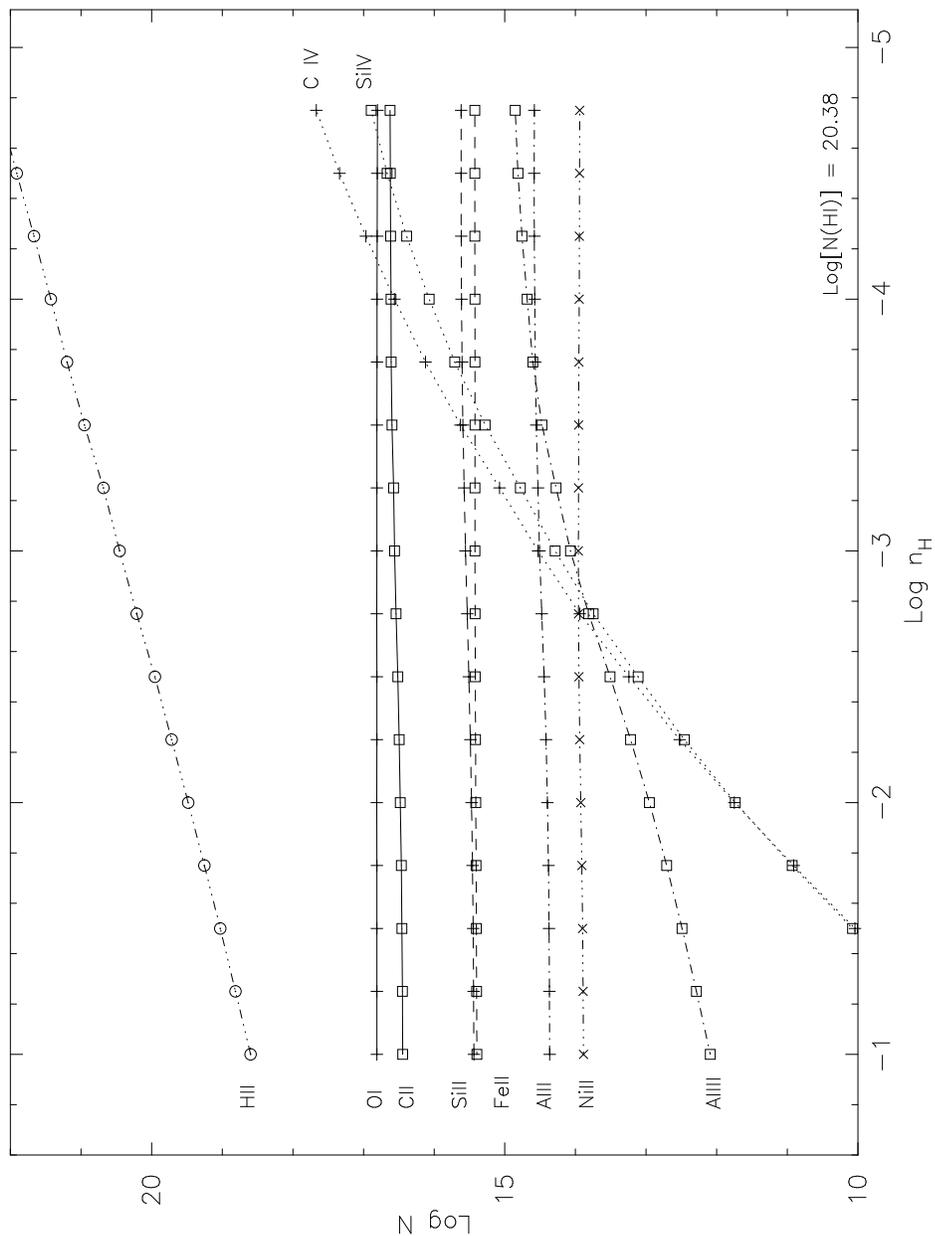,height=7.5in}}
\caption{ CLOUDY simulations for a series of systems with 
varying hydrogen volume
density assuming $[Z/H] = -0.5$ and requiring log[$\N{HI}$] = 20.38.  
The ionizing
spectra is an attenuated power law spectrum, calculated by Madau (1992) 
at an average
redshift of $z$=2.46. }
\label{CLOUDY}
\end{figure}

\begin{figure}
\centerline{
\psfig{figure=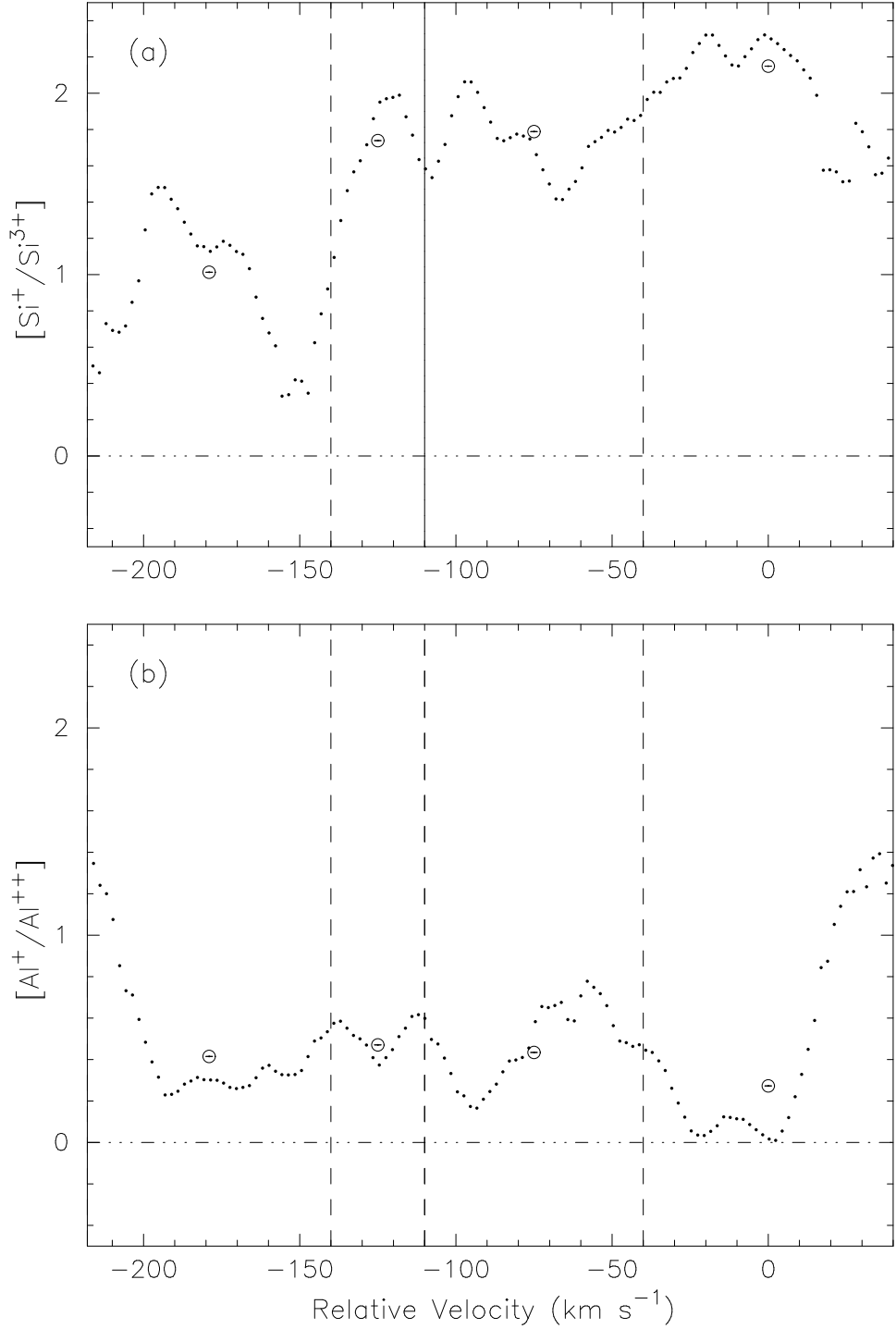,height=7.5in}}
\caption{Plot of the relative logarithmic column densities
of (a) Si$^+$ vs. Si$^{3+}$ and (b) Al$^+$ vs. Al$^{++}$.
The [Al$^+$/Al$^{++}$] ratio was derived assuming $\log \N{Al^+} = 12.0$
at each pixel and should be considered a lower limit.  The two
figures combined with our CLOUDY simulations further suggests that this
system is predominantly neutral.}
\label{Rtio}
\end{figure}

\begin{figure}
\centerline{
\psfig{figure=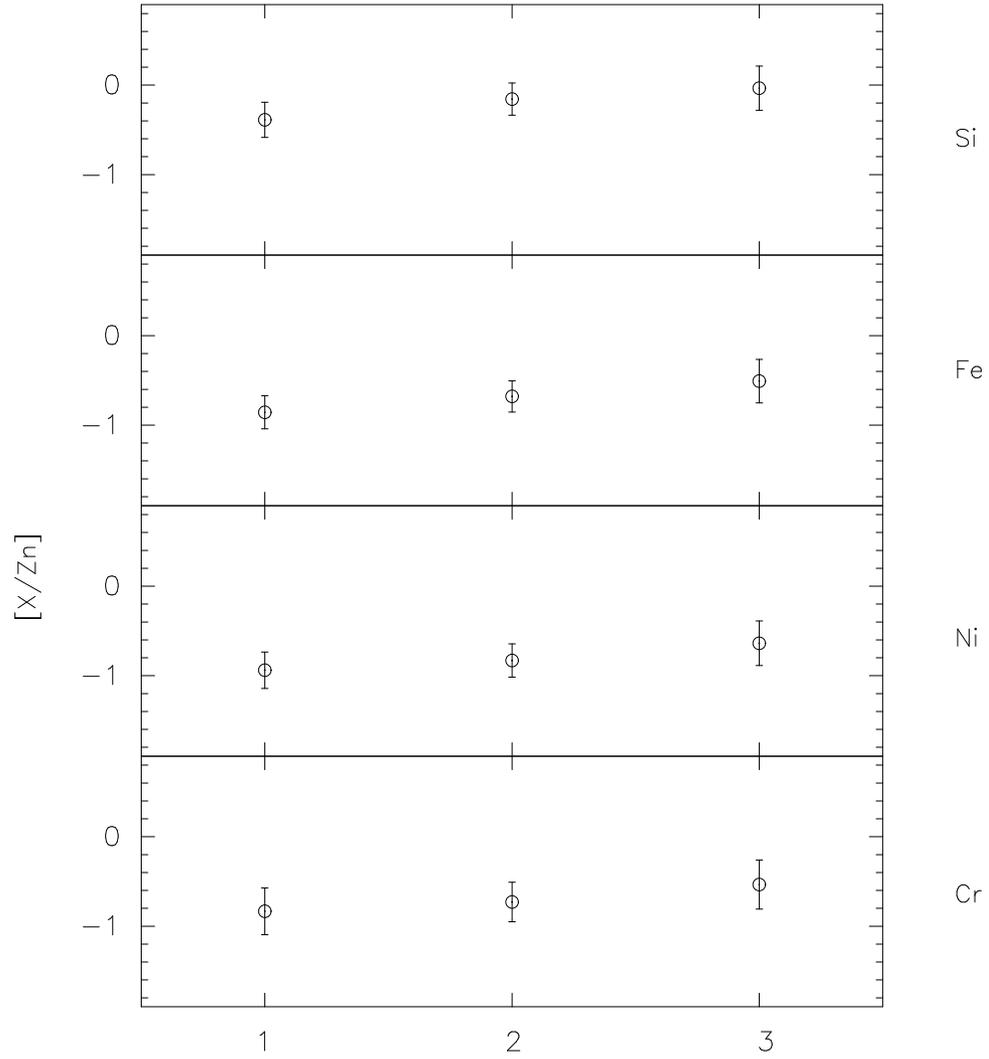,height=7.5in}}
\caption{ Relative cosmic abundances of Zn, Fe, Ni, Si, and Cr versus Zn for 3
velocity features of the $z$=2.462 system.  Note how the relative abundances are 
essentially the same
over all 3 features.}
\label{RelAbd}
\end{figure}

\begin{figure}
\centerline{
\psfig{figure=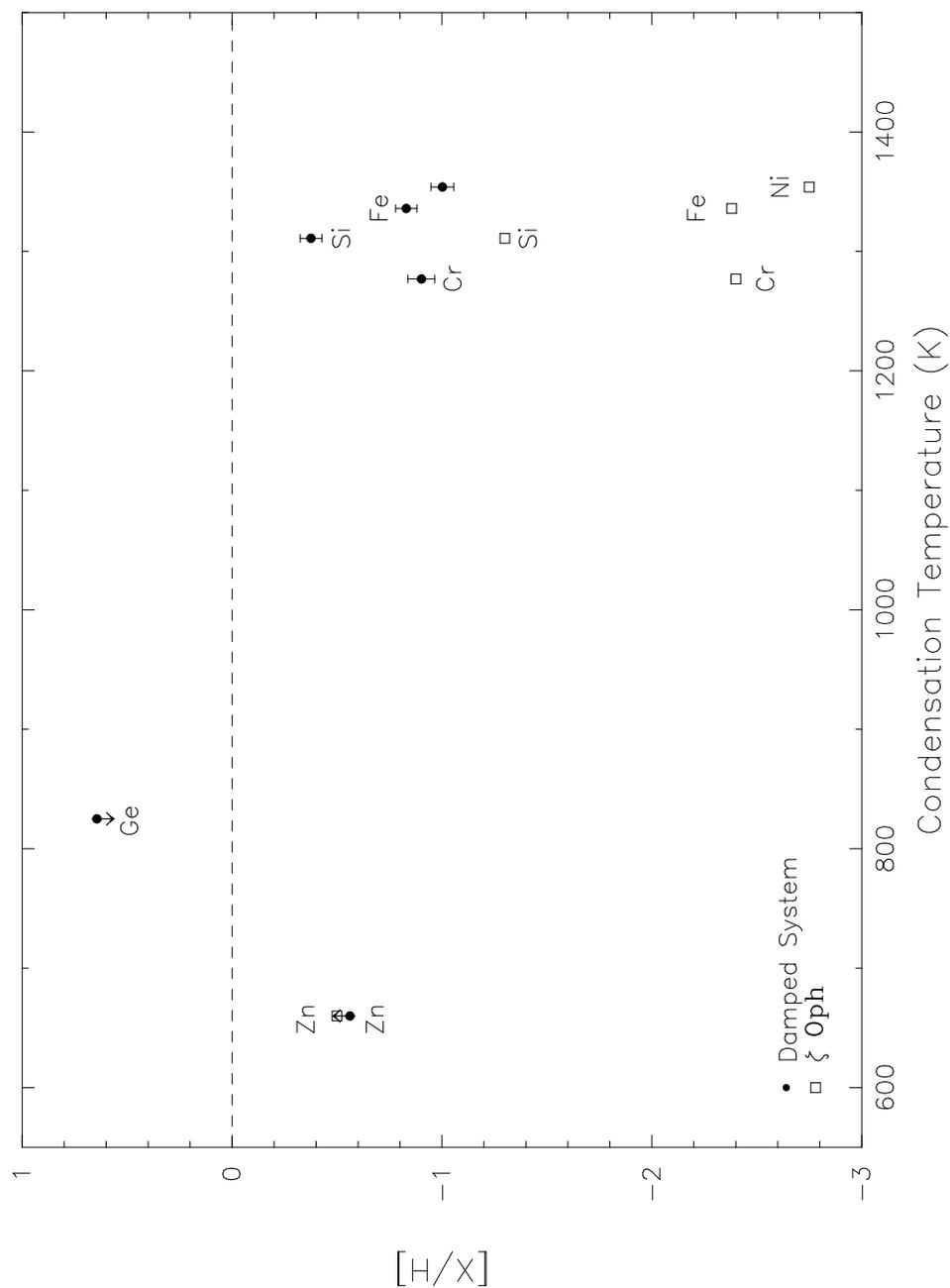,height=7.5in}}
\caption{ Logarithmic abundances of the low-ions
Fe, Cr, Si, Ni, Zn relative to standard
solar abundances versus condensation
temperatures for the $z$=2.462 system (solid dots) and for the line of sight
in the ISM toward $\zeta$ Oph (open squares). Note that the Zn abundance for
the damped \Lya system is a lower limit whose value is most likely
$\approx -0.26$.}
\label{Tcond}
\end{figure}

\begin{figure}
\centerline{
\psfig{figure=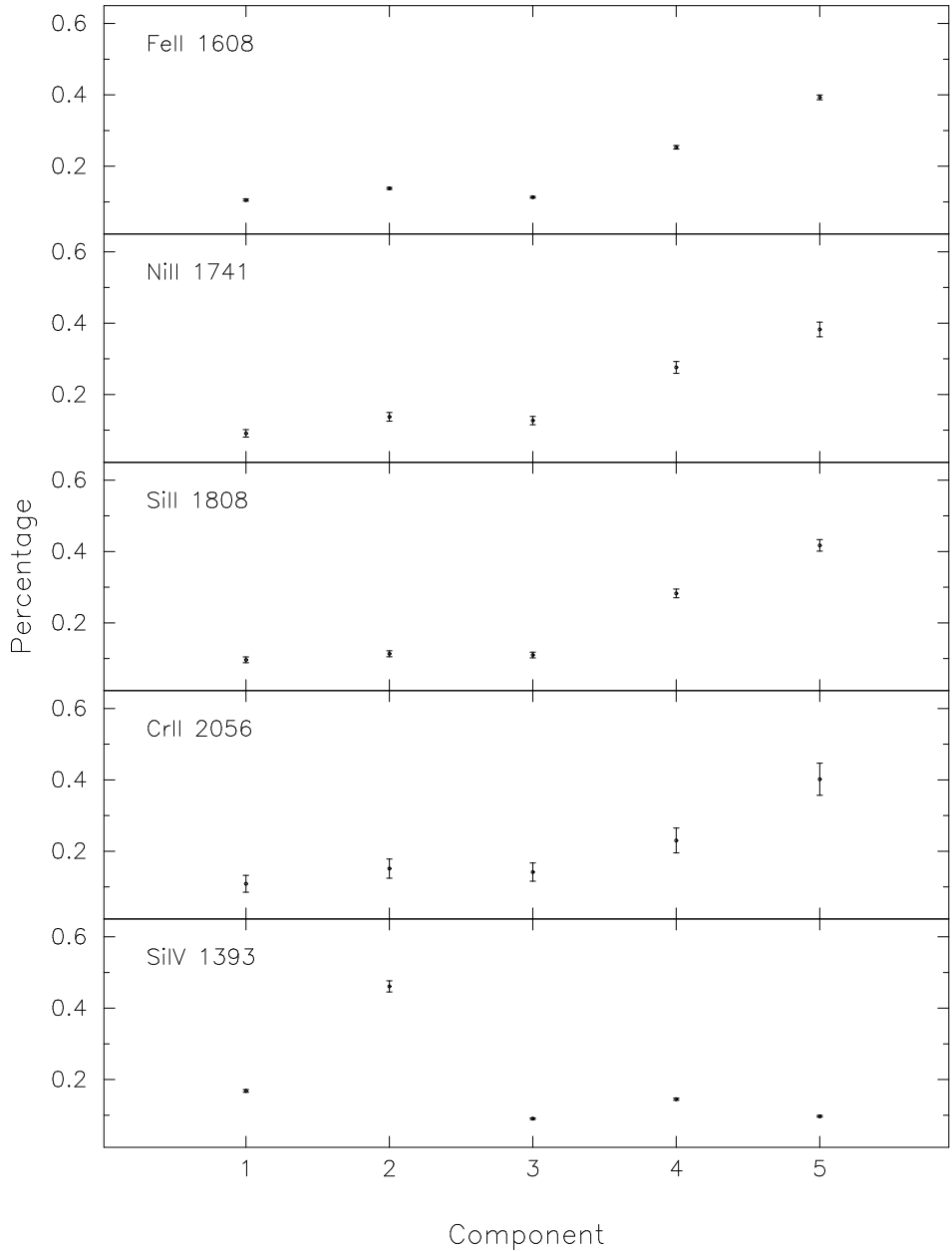,height=7.5in}}
\caption{ Percentage of total column densities in 5 features of 
the $z$=2.462 system
for the Fe II, Ni II, Si II, Cr II, and Si IV transitions.
Note the obvious differences between
the low and high-ions.
The errors were derived from the apparent optical depth method and
therefore should be taken only as an approximation.}
\label{Per-fig}
\end{figure}

\end{document}